\documentclass[fleqn,usenatbib]{mnras}

\usepackage{graphicx}
\usepackage{mathptmx}
\usepackage{savesym}
\usepackage{amsmath}
\savesymbol{iint}
\usepackage{txfonts}
\usepackage{lscape}
\restoresymbol{TXF}{iint}
\usepackage{tablefootnote}
\usepackage[T1]{fontenc}
\usepackage{ae,aecompl}
\usepackage{ragged2e}

\usepackage{amssymb}	
\usepackage{gensymb}
\usepackage{longtable}
\usepackage{lscape}
\usepackage{xcolor}

\title[Stellar evolution in NGC~663]{NGC~663 as a laboratory for massive star evolution \thanks{Partially based on observations collected at the Nordic Optical Telescope (La Palma), the Isaac Newton Telescope (La Palma) and the Observatoire de Haute-Provence.}}

\author[A. Marco et al.]{
Amparo Marco$^{1,2}$\thanks{E-mail: amparo.marco@ua.es (AM)}, 
Ignacio Negueruela$^{3,2}$, 
Norberto Castro$^{4}$ and
Sergio Sim\'on-D\'{\i}az$^{5,6}$
\\\\
$^{1}$DFISTS, Universidad de Alicante, Carretera San Vicente del Raspeig s/n,  E-03690, San Vicente del Raspeig, Spain\\
$^{2}$Instituto Universitario de Investigación Informática. Universidad de Alicante\\
$^{3}$Departamento de F\'{\i}sica Aplicada, Universidad de Alicante,Carretera San Vicente del Raspeig s/n,  E-03690, San Vicente del Raspeig, Spain \\
 $^{4}$ Leibniz-Institut für Astrophysik Potsdam (AIP), An der Sternwarte 16, D-14482, Potsdam, Germany\\
$^{5}$Instituto de Astrof\'{\i}sica de Canarias, E-38200, La Laguna, Tenerife, Spain\\
$^{6}$Departamento de Astrof\'{\i}sica, Universidad de la Laguna, Tenerife, Spain\\
}

\date{Accepted XXX. Received YYY; in original form ZZZ}

\pubyear{2024}

\begin{document}
\label{firstpage}
\pagerange{\pageref{firstpage}--\pageref{lastpage}}
\maketitle

\begin{abstract}
Massive young clusters with rich populations of high-mass stars are ideal laboratories to explore their evolutionary paths. Despite being the most prominent cluster in the Perseus-arm Cas~OB8 association, NGC~663 remains  comparatively little studied. We present a comprehensive investigation of its properties, integrating astrometric, photometric and spectroscopic data for the cluster and its surroundings, including accurate spectral classification for over 150 members. \textit{Gaia} astrometry indicates over 300 B-type members, possibly rendering NGC~663 the most massive cluster in the Perseus arm, with initial mass likely exceeding $10^4\:\mathrm{M}_{\sun}$. This large population makes NGC~663 an excellent laboratory for studying massive star evolution. Spectral analysis of the earliest members reveals approximately solar metallicity and a turn-off mass of $\approx8.5\:\mathrm{M}_{\sun}$, consistent with the photometric age of 23~Ma. We identify five spectroscopic blue stragglers, including the Be/X-ray binary RX~J0146.9+6121. We outline its evolutionary history and compare its properties with other Be stars. Although the cluster contains many Be stars, their relative fraction is not particularly high. Intriguingly, four of the six blue supergiant members appear to have significantly higher masses than the brightest giants near the Hertzsprung gap. These observations suggest that most mid-B supergiants may form via mergers, unless stars of 10\,--\,$12\:$M$_{\sun}$ born as primaries in binaries rarely undergo supernova explosions. Similarly, if Be stars form through the binary channel, then either most are produced through case A evolution or supernovae are uncommon among primaries in this mass range.

\end{abstract}




\section{Introduction}
The study of massive star evolution is subject to numerous uncertainties, many linked to stellar rotation and binary interaction \citep[e.g.][]{langer2012}. Massive young open clusters with a large and well characterized population of stars are the ideal laboratories to improve our understanding of these topics. The upper main sequences of populous young clusters have been known for long to contain stars that fitted poorly their turn-offs \citep[e.g.][]{marco2001}. Suspicions regarding the connection between these anomalous positions and binary evolution have long been held \citep[e.g.][]{marco2007} and have now solidified into certainties \citep[e.g.][and references therein]{bodensteiner2020}. Since the cluster is expected to have been formed at a given time (and so have a given age), stars that apparently deviate from the bulk of the population provide very valuable insights into the consequences of binary interactions. A key factor to exploit this capability is the correct determination of cluster membership. 

Prior to the advent of \textit{Gaia}, effects such as differential reddening rendered membership assessment for young clusters -- which were generally too far away for accurate proper motions -- very difficult. Any cluster member that did not fit the expected evolutionary sequence could be dismissed as a line-of-sight interloper. In the past few years, however,  \textit{Gaia} data have enabled precise determinations of membership for clusters at distances up to $\sim4$~kpc (except for very highly reddened systems; see \citealt{negueruela2022} for an example), by using the parallax and proper motion values \citep[e.g.][]{Cantat2020, hunt2023}. With this new situation, it is possible to explore the membership of stars belonging to the upper main sequence of the cluster, but occupying an unexpected position in its colour-magnitude diagram (CMD), as they will be key to understand anomalous evolutionary paths, while the bulk of the population will allow derivation of cluster parameters. After membership determination, we can use photometry and spectroscopy to make a detailed study of the cluster population. Paradoxically, such investigations have become commonplace for young massive clusters in the Magellanic Clouds \citep[e.g.][]{evans2006,bodensteiner2020}, where membership issues persist, given the depth of the galaxies, while many analogous clusters in our own Milky Way remain poorly explored. In this paper, we present a comprehensive study of one such cluster, NGC~663.

NGC~663 is one of the most massive open clusters in the Perseus arm, and the optimal environment to place in context a diverse array of evolved stars. This includes a significant number of blue supergiants (BSGs), several dozen Be stars, at least one red supergiant (RSG) and the only Galactic X-Ray binary that is associated with an open cluster, namely the Be/X-ray binary RX~J0146.9+6121 (with counterpart LS~I$+61\degree$235). Moreover, NGC~663 is the most massive of a number of open clusters sharing similar astrometric and physical characteristics that make up the historical Cas~OB8 association \citep{humphreys1978}.

Numerous studies of NGC 663 were conducted prior to the \textit{Gaia} era. Early investigations, based on photometric and spectroscopic studies  \citep[e.g.][]{hoag1961,moffat1972,bergh1978,svoloupoulos1962,hiltner1956,hoag1965}, yielded  distance moduli in the interval $11.5-12$, corresponding to distances between 2 and~2.5~kpc, and an age of $\sim10^7$~a. \citet{tapia1991} utilised near-infrared and Str\"{o}mgren photometry to obtain a distance modulus of $12.03\pm0.20$ ($\approx$2.5~kpc) and an age of $9\times10^6$~a. However, these age determinations pre-date the use of models with core overshooting and cannot be directly compared to modern values. More recently, \citet{pigulski2001}, by using $BVRI$ CCD photometry, adopted 11.6~mag as the true distance modulus of NGC~663 (equivalent to a distance of 2.1~kpc) and an age range between 20 and 25~Ma. \citet{pandey2005} examined the stellar contents of NGC~663 and the nearby -- and certainly related -- NGC~654 by using CCD $UBVI_{C}$ data and obtained a distance of $2.42\pm0.12$~kpc. In the current century, research focus has largely centred on identifying the Be star population of the cluster -- which was known to be substantial -- and characterizing its behaviour \citep{pigulski2001,yu2015,granada2018,dimitrov2018}.  

Astrometric data (positions, proper motions and parallaxes) from the European Space Agency (ESA) \textit{Gaia} mission have been used by a number of recent authors to determine cluster parameters for large collections of open clusters by using automated techniques for member determination and isochrone fitting in the $G$ vs. $(BP-RP)$ CMD. For NGC~663, \citet{Cantat2020}, by using \textit{Gaia} EDR2 data, obtained a distance of 2950~pc and an age $\log\,t= 7.47$ ($\approx 30$~Ma); \citet{dias2021}, with the same data, determined a distance of 2353$\pm$148~pc and an age $\log\,t= 7.40\pm0.25$ ($\approx 25$~Ma); finally, \citet{hunt2023}, with EDR3 data, resolved a distance of 2666$\pm$6~pc and an age $\log\,t= 7.44\pm0.19$ ($\approx 28$~Ma). Observing the different values of the distance modulus and ages determined in previous works, we can say that the age has always been around 25~Ma, but the distance derived was in all cases shorter than the value given by \textit{Gaia} data, a surprising fact given the degeneracy between age and distance that plagued older work. Perhaps because of this underestimation in the distance modulus, sometimes by over a magnitude, NGC~663 has never been considered a high-mass cluster or of particular relevance for the study of early-type stars (beyond the Be phenomenon) despite the six BSGs that have traditionally been associated with it --  a number that exceeds those associated with any other Perseus arm cluster, unless we consider the whole association surrounding the double cluster $h$ \& $\chi$~Per.

Likewise, the Cas~OB8 association is very poorly studied. According to 
\citet{humphreys1978}, this association is formed by a number of supergiants spread between 
some clusters with approximately the same age, i.e.\ NGC~663, NGC~654, NGC~659 and NGC~581.
While NGC~663 is the largest cluster in the region and has a substantial number of supergiants in the core and the halo \citep{marco2007,pandey2005,mermilliod2008}, NGC~654, located about 30~pc North of NGC~663, is smaller and contains one blue and one yellow supergiant \citep{marco2007,pandey2005}. Another even smaller cluster, NGC~659, is $\approx25$ pc to the SE of NGC~663 and contains no supergiants. NGC~581 is further away, around $65$~pc to the East and contains a blue and a red supergiant.


In this paper, we concentrate on the main cluster NGC~663 and its halo, presenting Str\"{o}mgren photometry for its central region, and optical spectroscopy for 153 stars in the field, the vast majority being cluster members. With the analysis of this dataset and the \textit{Gaia} data, we are able to study with better accuracy the population of the cluster, obtaining spectral types and stellar parameters. In addition, we estimate the mass of the cluster, confirming that it is the most massive open cluster in the Perseus arm, except perhaps for NGC~7419. Our final scope is to investigate the evolution of the upper main sequence population, which is not possible by using only \textit{Gaia} data \citep[cf.][]{negueruela2023}.

The paper is divided following this structure: in Section~\ref{data}, we present the observations and reduction procedures used; in Section~\ref{results} we determine the parameters and the mass of the cluster and characterize its population by determining spectral types and stellar atmospheric parameters; in Section~\ref{discussion} we analyse the upper main sequence population, together with the theories of mechanism of binary star formation. Finally, we list the conclusions.       

\section{Observations and data}
\label{data}

We observed the large spectroscopic dataset at low-medium resolution with two different instruments, the Intermediate Dispersion Spectrograph (IDS) on the 2.54-m Isaac Newton Telescope (INT) at the La Palma observatory on the nights of 2002 October 25\,--\,28, and the long-slit spectrograph \textit{Carelec} on the 1.93-m telescope at the Observatoire de Haute-Provence (from now on, T1.9) on the nights 2002 November 1\,--\,8. High resolution spectra were taken with the FIbre-fed Echelle Spectrograph (FIES) on the 2.56-m Nordic Optical Telescope (NOT) at the La Palma observatory on the nights of 2011 January 11\,--\,15. 

We used the imager and spectrograph Andaluc\'{i}a Faint Object Spectrograph and Camera (ALFOSC), also on the NOT, to obtain Str\"omgren photometry on the night of 2005 October 3. Finally, we made use of \textit{Gaia} EDR3 data corresponding to the field of NGC~663 \citep{Gaia Collaboration3}.

\subsection{Optical photometry}
\label{opt_phot}

ALFOSC allows observations in different modes. In imaging mode, the camera covers a field of $6\farcm5 \times 6\farcm5$ and has a pixel scale of $0\farcs19\:$pixel$^{-1}$. We took Str\"omgren photometry from different frames in order to cover the majority of the stars belonging to the cluster. 
For all frames, we obtained two series of different exposure times in each filter to achieve accurate photometry for a broad magnitude range (see Table~\ref{tab:tab1}). 
The standard stars observed and the reduction procedure for standard and target fields is the same used in section~2.1 of \citet{marco2013}, as the observations reported there were part of the same observing run.

We obtained Str\"{o}mgren photometry for 289 stars in the field of NGC~663. The number of stars detected in all filters is limited by the long exposure time in the $u$ filter. From these, only 180 are \textit{Gaia} members. In Table~\ref{tab:membersphot} we list the photometry for these 180 stars, their coordinates in J2000, their values of $V$, $(b-y)$, $m_{1}$, $c_{1}$ and $\beta$ with the corresponding errors (the standard deviation when there are more than one measurement for each value, or the addition in quadrature of the contribution to the total error made by each photometric individual error given by {\sc daophot} in the opposite case) and the number of measurements for each magnitude or colour. If a spectrum of the star has been taken, this is also indicated in the table. Astrometric referencing of all our images was made by using the same procedure as in \citet{marco2016}.

\begin{table}
	\centering
	\caption{Log of the photometric observations taken at the NOT in October 2005.\label{tab:tab1}}
        \begin{tabular}{ccc}
        &Exposure time (s) &\\
       Filter&Long times&Short times\\
       \noalign{\smallskip}
        \hline
         \noalign{\smallskip}
        $u$&30&10\\
        $v$&12&5\\
        $b$&10&3\\
        $y$&5&2\\
         \noalign{\smallskip}
        \hline
        \end{tabular}
\end{table}

 \subsection{Spectroscopy}

The main spectroscopic dataset was observed with the IDS spectrograph, equipped with the 235-mm camera and the EEV10 CCD. We used the R1200B grating for the brightest members (covering  3900\,--\,5200\AA\,) and the R400V  grating  for  fainter  stars (covering 3500\,--\,7200\AA\,). In both cases, we set a $1.2 \arcsec$ slit width, obtaining a resolving power $R\sim4000$ for R1200B and $R\sim1400$ for R400V. Because the main aim was to derive spectral classification, we did not take a sufficient number of arcs to measure radial velocities. The lists of stars observed with IDS at $R\sim4000$ and $R\sim1400$ can be seen in Table~\ref{tab:R4000} and Table~\ref{tab:R1400}, respectively. A representative sample of stars observed with the R1200B grating can be seen in Fig.~\ref{fig:giants}. A similar sample for the R400V grating can be found in the appendix (Fig.~\ref{R400sample}).

\begin{figure*}
\centering
\resizebox{0.48\textwidth}{!}{\includegraphics[]{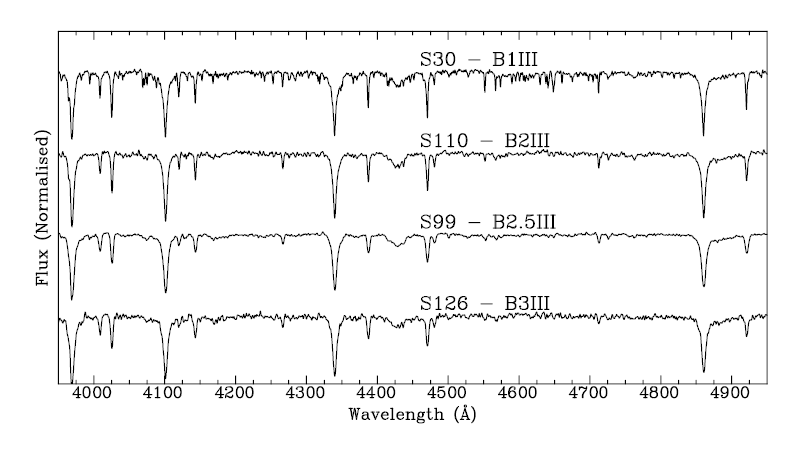}}
\resizebox{0.48\textwidth}{!}{\includegraphics[]{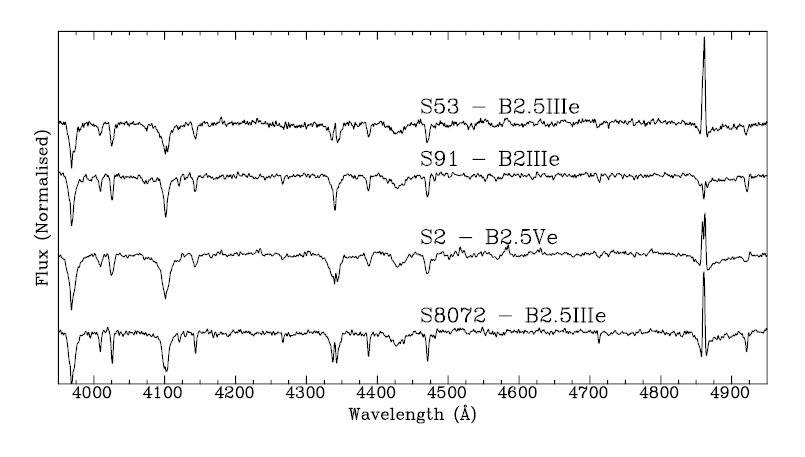}}
\caption{Some bright members of NGC~663 observed with IDS + R1200B. The spectra have $R\approx4\,000$ and varying SNR, in almost all cases > 100. The \textit{left panel} displays some of the most luminous giants without emission lines. The \textit{right panel} shows a sample of Be stars. At this resolution and SNR, spectral classification is not seriously impaired by the presence of emission lines, even in cases with a significant number of Fe\,{\sc ii} lines in emission, such as S2 and S53. \label{fig:giants}}
\end{figure*} 

This set was complemented by the T1.9 observations, at lower resolution. The  \textit{Carelec} long-slit spectrograph was equipped with the EEV42-20 CCD and the 300~ln$\:$mm$^{-1}$ grating, covering the 3800\,--\,6800\AA\, spectral range. With a 2$\arcsec$ slit width (suited to the local seeing), we obtained a resolving power $R\sim900$. The list of stars observed with \textit{Carelec} can be found in Table~\ref{tab:R900}, while a representative sample of stars can be found in the appendix (Fig.~\ref{OHPsample}).

All the long-slit spectroscopic data were reduced  with  the \textit{Starlink} packages {\sc CCDPACK} \citep{Draper2000} and {\sc FIGARO} \citep{Shortridge1997}, and analysed using {\sc FIGARO} and {\sc DIPSO} \citep{Howarth1998}. 

A collection of stars classified as BSGs in and around the cluster were observed at higher resolution with FIES. We selected the low-resolution mode, with a resolving power $R\sim25\,000$. At the time, FIES was equipped with CCD13, resulting in coverage over the 3700\,--\,7300\AA\, spectral range without any gaps \citep{Telting2014}. The FIES spectra were homogeneously reduced using the FIEStool{\footnote{http://www.not.iac.es/instruments/fies/fiestool/FIEStool.html}} software in advanced mode. Every night, a complete set of bias, flat, and arc frames was taken, while wavelength calibration was accomplished with arc spectra of a ThAr lamp. The list of stars observed with FIES is given in Table~\ref{tab:SGFIES}, while the spectra are displayed in Fig.~\ref{fig:sgs}.

\begin{table*}
	\centering
	\caption{Basic parameters for blue supergiant stars in the area of NGC~663 that were observed with FIES. The spectral types are from this work, the IDs are taken from the WEBDA database, and the astrometry corresponds to \textit{Gaia} EDR3. Parallaxes have been zero point corrected. Stars in the top panel are considered astrometric members. Stars in the bottom panel are astrometric non-members, but the parameters of BD~$+60\degree$\,336 are compatible with having been ejected from the cluster core. Errors in derived parameters are dominated by the grid step and can be taken as $\pm1000\:$K in $T_{\mathrm{eff}}$ and $\pm0.1$~dex in $\log\,g$. The corresponding spectra are displayed in Fig.~\ref{fig:sgs} and example model fits are shown in Fig.~\ref{fig:fits}. \label{tab:SGFIES}}

\begin{tabular}{cclllcccc} 
\hline
\noalign{\smallskip}
RA&DEC&ID&Name&Spectral&$G$&Plx&pmRA&pmDE\\
(J2000)&(J2000)&&&Type&&(mas)&(mas$\cdot\mathrm{a}^{-1}$)&(mas$\cdot\mathrm{a}^{-1}$)\\
            \noalign{\smallskip}
            \hline
            \multicolumn{9}{c}{\textbf{Members}}\\
\noalign{\smallskip}
\hline
\noalign{\smallskip}
01:46:34.53&$+$61:15:45.0&830&BD~$+60\degree$\,343&B2.5\,Ib&9.1 &0.374$\pm$0.014&$-$1.164$\pm$0.009&$-$0.318$\pm$0.010\\
\noalign{\smallskip}
01:47:20.16&$+$61:07:54.8&221&BD~$+60\degree$\,351&B3\,Ib&8.9& 0.361$\pm$0.014&$-$1.076$\pm$0.010&$-$0.413$\pm$0.011\\
\noalign{\smallskip}
01:46:04.88&$+$61:13:41.9&54&BD~$+60\degree$\,333&B4\,Ib&8.7 &0.369$\pm$0.017&$-$0.964$\pm$0.011&$-$0.356$\pm$0.013\\
\noalign{\smallskip}
01:46:23.77&$+$61:15:29.8 &44&BD~$+60\degree$\,339&B5\,Ib&8.3&0.385$\pm$0.016&$-$1.124$\pm$0.011&$-$0.345$\pm$0.013\\
\noalign{\smallskip}
01:45:56.10&$+$61:14:05.5&86&BD~$+60\degree$\,331&B8\,Ib&8.6 & 0.379$\pm$0.014&$-$1.105$\pm$0.010&$-$0.372$\pm$0.011\\
\noalign{\smallskip}
01:47:08.89&$+$61:11:54.3&147&BD~$+60\degree$\,347&B9\,Ib&9.2 & 0.356$\pm$0.013&$-$1.121$\pm$0.009&$-$0.399$\pm$0.010\\
\noalign{\smallskip}
\hline
\multicolumn{9}{c}{\textbf{Non-members}}\\
\hline
\noalign{\smallskip}
01:42:58.32& $+$61:25:15.9&&HD~10362&B7\,II &6.3 & 2.451$\pm$0.033&2.525$\pm$0.024&-9.154$\pm$0.023\\
\noalign{\smallskip}
01:46:56.38&$+$60:40:11.2&&HD~10756&B8\,Iab-Ib&7.4& 0.318$\pm$0.030&$-$1.011$\pm$0.016&$+$0.136$\pm$0.018\\
\noalign{\smallskip}
01:46:10.20&$+$61:26:33.1&307&BD~$+60\degree$\,336&A0\,Ib&8.8 &0.395$\pm$0.013&$-$1.241$\pm$0.009&$+$0.177$\pm$0.011\\
\noalign{\smallskip}
01:52:58.69&$+$60:51:15.43& & BD~$+60\degree$\,369& B2\,II & 9.4 &0.377$\pm$0.013&$-$0.782$\pm$0.009&$-$0.988$\pm$0.010\\
\noalign{\smallskip}
\hline
\end{tabular}
\end{table*}

\section{Results}
\label{results}
		  
\subsection{Spectral classification}

Spectral classification was carried out by comparing our spectra to the new grid of MK standard stars presented in \citet{negueruela24}. For the INT sample, resolution is sufficient to determine luminosity class by comparing the wings of Balmer lines, and the spectral type was then determined by application of the classification criteria defined by these authors. For the T1.9 sample, spectral type was determined by direct comparison of the overall spectrum. Internal consistency was achieved by multiple comparisons to other sample stars with similar spectral types. Many of the stars observed at $R=4000$ were also observed at $R=1400$, and these spectra were used as comparators for the rest of the lower resolution spectra. Again, a number of stars were observed with both the INT and the T1.9 and used to guarantee internal consistency. Nevertheless, the spectral types derived from T1.9 spectra must be considered less accurate than those coming from INT spectra, and we only claim an accuracy of $\pm1$ spectral type and luminosity class (although, given their position in the CMD, the vast majority of the objects must have luminosity class V).


As noted, the cluster contains a very significant number of Be stars. The spectral classification of Be stars may be complicated because of the presence of many emission lines mixed with the photospheric absorption profiles \citep[e.g.][]{steele1999}, mostly at lower resolution. As can be seen in Fig.~\ref{fig:giants} (right panel), the emission lines do not affect significantly classification criteria at $R=4\,000$. 

\begin{figure*}
\centering
\resizebox{\columnwidth}{!}{\includegraphics[clip]{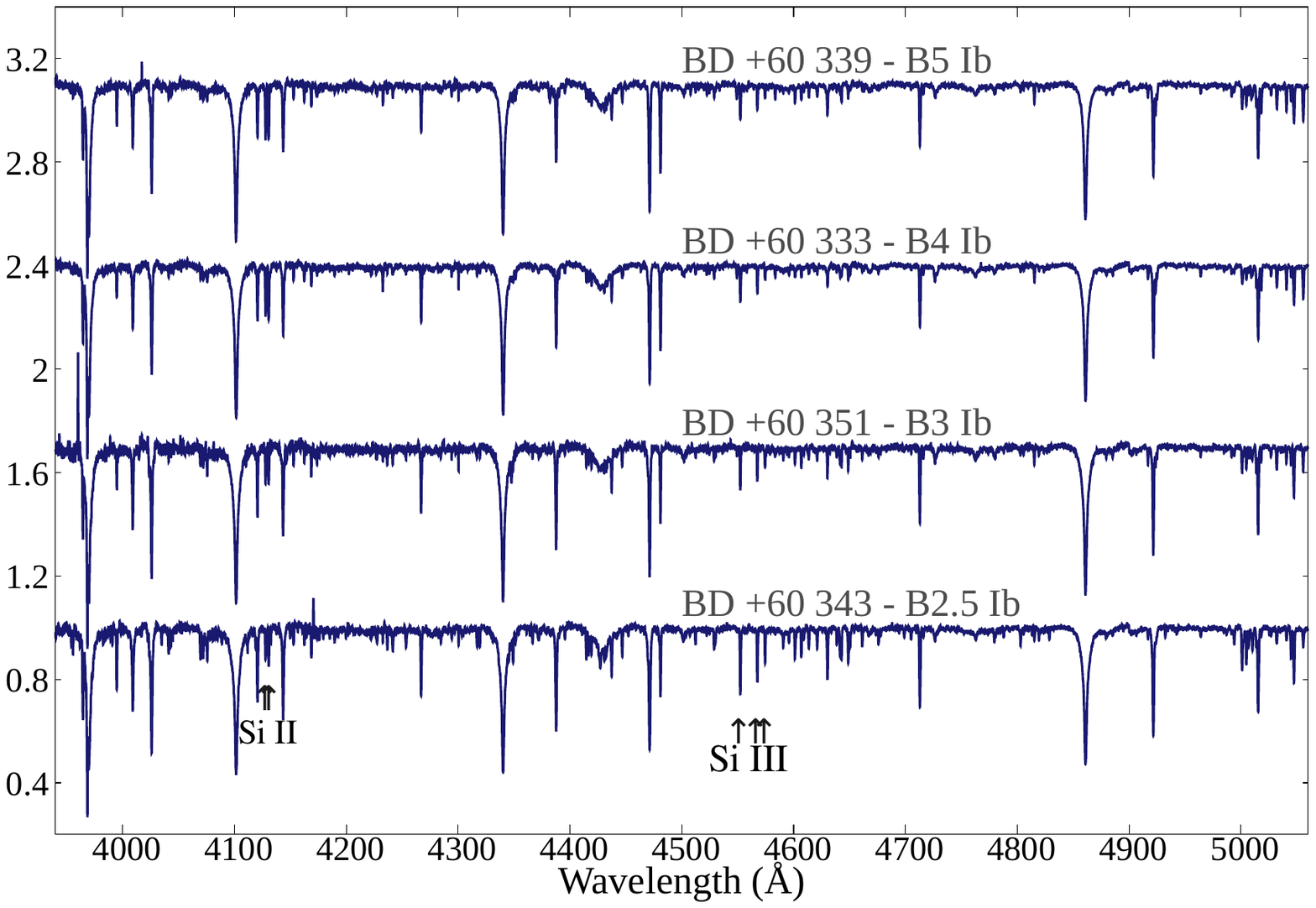}
}
\resizebox{\columnwidth}{!}{\includegraphics[clip]{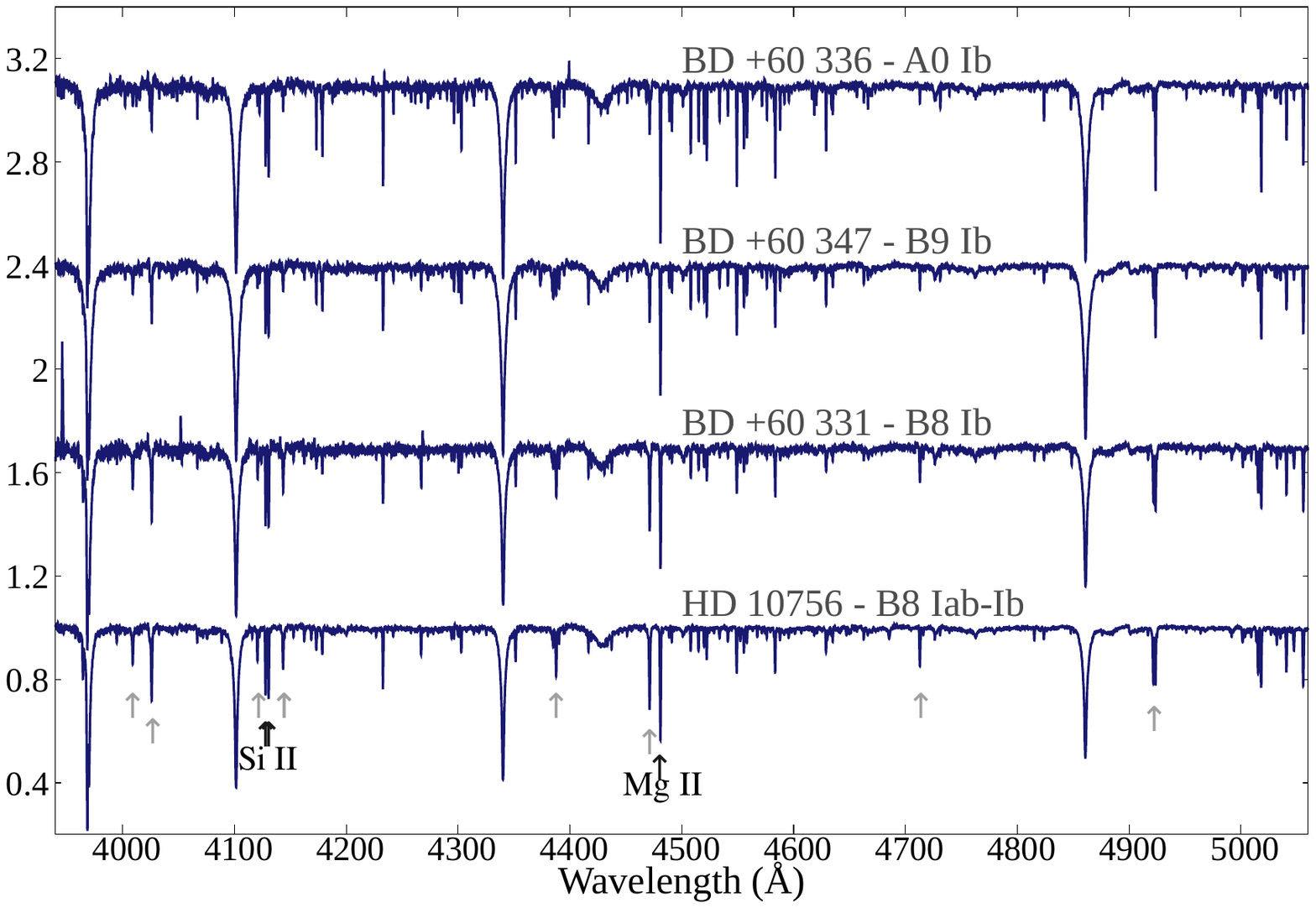}
}
\caption{High-resolution (FIES) spectra of blue supergiants in the area of NGC~663. \textit{Left}: Mid-B supergiants: note the increase in the \ion{Si}{ii}/\ion{Si}{iii} ratio as we move to later spectral types.  \textit{Right}: Late-B supergiants. At approximately constant luminosity, \ion{Si}{ii} and \ion{Mg}{ii} increase as we move to later types, while \ion{He}{i} lines (marked with light-grey arrows) weaken until they virtually disappear in the A-type stars\label{fig:sgs}.}
\end{figure*}

\subsection{Stellar atmospheric parameters}
\label{sec:steparams}

We derived the stellar atmospheric parameters using the method described by \citet{castro2012}, which involves an automated, $\chi^2$-based algorithm that identifies the model spectrum that best reproduces key hydrogen, helium, and silicon lines in the 4000\,--\,5000\,$\AA$ range. The specific lines employed in the analysis are listed in table~1 of \citet{castro2012}. Their dependence on effective temperature ($T_{\mathrm{eff}}$) is illustrated in Fig.~\ref{fig:sgs}. The strength of the \ion{Mg}{ii}~4481\,\AA\ line is also considered by the fitting algorithm, as this may be the strongest metallic line in late-B stars.

We applied this method to three of our datasets: the FIES spectra of BSGs, and the INT spectra of bright members obtained at both $R=4\,000$ and $R=1\,400$. We did not include spectra with low SNR or spectra of Be stars with strong emission lines (for example, S53 or S2 in the right panel of Fig.~\ref{fig:giants}, where the wings of the photospheric H$\beta$ line cannot be seen). We did include Be stars exhibiting weaker emission and narrow photospheric lines, such as S91 and S8072 in Fig.~\ref{fig:giants} (right panel), although their parameters should be regarded as less reliable. 

The analysis was primarily performed using grids of synthetic stellar atmosphere models generated with the {\sc fastwind} code \citep{santolaya1997,puls2005}, as detailed in \citet{castro2012}. However, several stars in the INT sample exhibit surface gravities higher than those covered by those original grids, which were initially designed for giant and supergiant B stars. For these stars, which lie near or beyond the upper gravity limit or at higher temperatures than the supergiant grids, we conducted a second iteration of the analysis using an auxiliary grid of models developed specifically for this project. This supplementary grid extends to higher gravities and temperatures and assumes solar abundances for all the elements.

The auxiliary grid uses the same atomic models as the original supergiant grids, allowing the same diagnostic lines to be used consistently throughout the analysis. The grid is designed to include those combinations of $T_{\mathrm{eff}}$ and $\log\,g$
that main sequence and giant OB stars can realistically be
expected to have, covering values of $\log\,g$ between 2.0 and~4.2 and $4.1\leq \log T_{\mathrm{eff}}\leq4.8$.


We used the IACOB-BROAD code \citep{simon14} to measure the projected rotational velocity ($v \sin\, i$) and macroturbulent velocity ($v_\mathrm{mac}$) for stars having the higher resolution FIES spectra. This tool characterises the additional broadening in stellar line profiles by combining Fourier transform techniques with a goodness-of-fit approach.

For stars observed at lower resolution with the IDS spectrograph, we initially adopted a default $v \sin\,i$ of $50\:$km\,s$^{-1}$, assuming that the observed line broadening is dominated by the instrumental profile. We then compared the best-fit model, obtained using the $\chi^2$-based algorithm, with the observed spectrum. If the instrumental broadening was insufficient to reproduce the observed line profiles, we estimated a revised $v \sin\,i$ and repeated the analysis. This process was iterated until no significant improvement was found by further increasing the rotational velocity. Given the low resolution of the spectra, especially the $R=1400$ sample, this should be taken as a rough estimate of the presence of significant rotational broadening, and not an actual measurement.


The formal errors in parameters reflect the number of models that can be considered compatible with the observed spectrum, and must be at least as large as the grid step. The parameters derived from the FIES spectra are displayed in Table~\ref{tab:SGpar}, while those found from the analysis of lower-resolution spectra are listed in Table~\ref{tab:analysis}. 

Considering the low resolution of the spectra used and the moderate SNR in some cases, the parameters derived for non-emission stars are to a very large degree consistent with the spectral classifications. The bright B2\,--\,B2.5 giants at the top of the sequence all have $\log\,g$ close to 3.1, while stars classified as IV have $\log\,g\approx3.7$ and main sequence stars have $\log\,g\approx4.0$. The fits for the mild Be stars are considerably worse. The effective gravities are in many cases high for the spectral type (for instance, in all the giants) and the temperatures can thus be inaccurate. In particular, three Be stars -- namely 21, 91 and 132 -- have a $T_{\mathrm{eff}}$ far too hot for their spectral type. Consequently, they will not be used in the discussion.


The six supergiants that are astrometric members of the cluster were observed both with FIES and the IDS at $R=4\,000$. The parameters derived are in all cases compatible within the relatively large errors of the IDS data, and therefore we will adopt those derived from the FIES sample. We tried excluding H$\beta$ from the fit, because its possible contamination by a diffuse insterstallar band, but found no difference in the effective gravity of the best model fits.


\begin{figure*}
\resizebox{\textwidth}{!}{\includegraphics[angle=0]{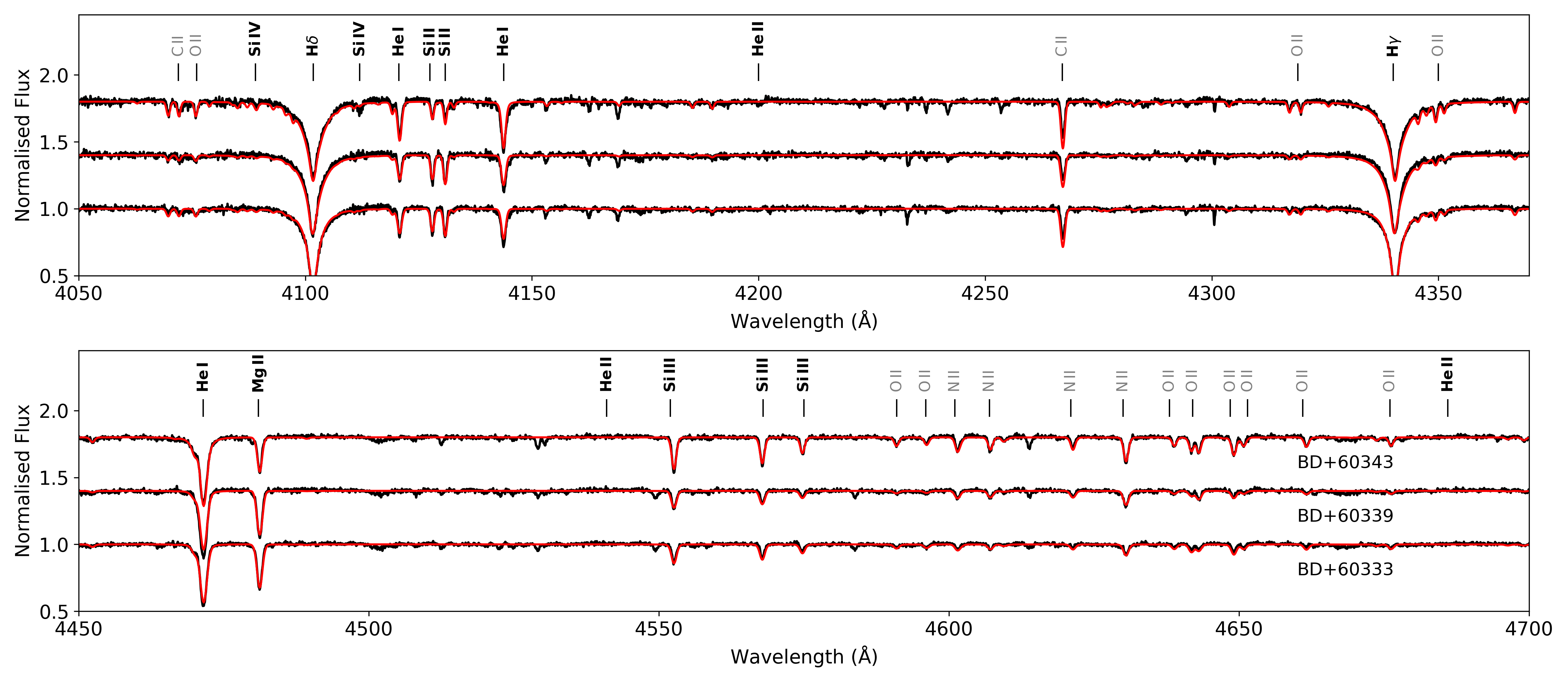}}
\centering
\caption{Model spectra (red) fitted to the echelle spectra of three of the supergiant members. \textbf{Top panel:} Region including the H$\delta$ and H$\gamma$ lines, whose wings determine the effective gravity. \textbf{Bottom panel:} Region containing the majority of metallic lines. The diagnostic lines used in the fit are marked on top of the spectrum of BD~$+60\degree$343 as black labels. C, N, O lines that are generally utilised for abundance determination are indicated with grey labels. \label{fig:fits}} 
\end{figure*}

Of the four astrometric non-members observed, BD~$+60\degree$\,336 is an A0\,Ib star to the North of NGC~663, whose proper motions suggest it has been ejected from the cluster. It is the coolest object in our sample and therefore affected by the same limitations as BD~$+60\degree$\,347. HD~10756 is a luminous supergiant to the South of the cluster. It is likely a member of Cas~OB8, although its proper motions are somewhat peculiar. HD~10362 turns out to be a foreground object of lower luminosity.

The recent study by \citet{deBurgos24} provides stellar parameters for a large sample of Galactic BSGs. Among them, we find the four hotter supergiants in our sample. Additionally, their sample includes two other stars in the area. BD~$+60\degree$\,345 (star 376) is a giant in the halo of the cluster, with astrometric parameters (listed in Table~\ref{tab:R1400}) fully compatible with membership, for which we only had a low-resolution spectrum. BD~$+60\degree$\,369 is a catalogued B-type supergiant located about $50\arcmin$ to the southeast of NGC~663. For completeness, we list its parameters, which are consistent with membership in Cas~OB8, at the bottom of Table~\ref{tab:SGFIES}. We have downloaded their spectra and added these two stars to our analysis, with the aim of having a larger sample to allow a comparison between the two works. The temperature determinations by \citet{deBurgos24} are based on the ionization balance between different Si ions, a slightly different method\footnote{The rotational velocity parameters ($v\,\sin\,i$, $v_{\mathrm{mac}}$) for all the stars hotter than 15\,000~K are identical in \citet{deBurgos24} and this work, having been calculated with the same tool on the same spectra.}. We find very good agreement for all the stars in common, except BD~$+60\degree$\,343 and BD~$+60\degree$\,369, for which the differences are slightly larger than our grid step in both $T_{\mathrm{eff}}$ and $\log\,g$. Although the proper motions of BD~$+60\degree$\,369 do not seem compatible with a direct ejection from the cluster core, its high rotational velocity (for a supergiant) and its divergent proper motions strongly point to a binary origin.

The parameters for the five stars with effective temperatures lower than 15\,000~K cannot be considered fully reliable\footnote{The version of the \textsc{fastwind} code used in this work does not include a complete and fully tested \ion{Fe}{ii} model atom for computing background opacities in the approximate line-blanketing treatment implemented in the code (J.~Puls, priv. comm.). This limitation is expected to affect the reliable determination of effective temperatures and surface gravities for stars with $T_{\mathrm{eff}}\la15\,000$~K.}. We include them for completeness, but they are not decisive for the  discussion that follows.

\begin{table*}
	\centering
	\caption{Stellar parameters for the blue supergiant stars, plus the blue straggler BD~$+60\degree$\,345, derived from their FIES spectra. The top panel lists certain members, while the bottom panel contains astrometric non-members. Errors are dominated by the grid step and can be taken as $\pm1000\:$K in $T_{\mathrm{eff}}$ and $\pm0.1$~dex in $\log\,g$. \label{tab:SGpar}}
\begin{tabular}{lccccc} 
	  \hline
           \noalign{\smallskip}
           Name&Spectral & $v\,\sin\,i$ & $v_\mathrm{mac}$& $T_{\mathrm{eff}}$&$\log\,g$\\
           & Type & (km\,s$^{-1}$)& (km\,s$^{-1}$)& (K)  & \\
	   \noalign{\smallskip}
\hline
\multicolumn{6}{c}{\textbf{Members}}\\
\noalign{\smallskip}
\hline
\noalign{\smallskip}
BD~$+60\degree$\,343&B2.5\,Ib& 30 & 33 &$18\,000\pm1000$&$2.7\pm0.1$\\
\noalign{\smallskip}
BD~$+60\degree$\,351&B3\,Ib &16 & 38 & $17\,000\pm1000$&$2.7\pm0.1$\\
\noalign{\smallskip}
BD~$+60\degree$\,339&B5\,Ib & 21 & 56 & $15\,000\pm1000$&$2.4\pm0.1$\\
\noalign{\smallskip}
BD~$+60\degree$\,333&B4\,Ib & 25& 50 & $15\,000\pm1000$&$2.3\pm0.1$\\
\noalign{\smallskip}
BD~$+60\degree$\,331&B8\,Ib& 29 & 43 & $12\,000\pm1000$&$2.0\pm0.3$\\
\noalign{\smallskip}
BD~$+60\degree$\,347&B9\,Ib& 29 & 30 & $11\,000\pm1000$&$2.3\pm0.1$\\
\noalign{\smallskip}
BD~$+60\degree$\,345&B0\,III&32 & 64 & $29\,000\pm1000$& $3.3\pm0.1$\\
\noalign{\smallskip}
\hline
\multicolumn{6}{c}{\textbf{Non-members}}\\
\hline
\noalign{\smallskip}
HD~10362&B7\,II & 34 & 9 & $12\,000\pm1000$&$2.8\pm0.1$\\
\noalign{\smallskip}
HD~10756&B8\,Iab--Ib& 29 & 36 & $12\,000\pm1000$&$2.0\pm0.1$\\
\noalign{\smallskip}
BD~$+60\degree$\,336&A0\,Ib& 25 & 23 & $10\,000\pm1000$&$1.9\pm0.1$\\
\noalign{\smallskip}
BD~$+60\degree$\,369&B2\,II& 94& 64 & $19\,000\pm1500$& $2.7\pm0.1$\\
\hline
\end{tabular}
\end{table*}

Although we did not try a full abundance determination, the spectra of all the BSGs are well reproduced by models of approximately solar composition, with some evidence for a slight C deficiency and N enhancement with respect to main sequence B stars \citep{nieva2012}, typical of such stars (see examples in Fig.~\ref{fig:fits}). It is also worth noting that all our spectra are well fit by models with typical He abundances. \citet{deBurgos24} derive He abundances in a more soffisticated way, finding that all stars are compatible with a standard 0.10 abundance, although in the case of BD~$+60\degree$\,369 the errors are very large. An approximately solar abundance is in agreement with measurements of metallicity from classical Cepheids, not only the value expected at the Galactocentric radius of NGC~663 according to the metallicity gradient \citep{kovtyukh2022,dasilva2023}, but also individual measurements of Cepheids in the cluster's vicinity, such as RW~Cas, AW~Cas or VV~Cas \citep[e.g.][]{luck2011}. Notwithstanding, \citet{fanelli2022} derived abundances for a large sample of RSGs in this region of the Perseus Arm, finding very subsolar values, similar to those of the LMC. The reasons for this huge discrepancy have not been explored. Given the high degree of consistency of the Cepheid values between different groups and sets of models, and our own results, we will use solar-metallicity isochrones throughout the paper.

\subsection{Analysis with Gaia data}
\label{sec:gaia}

Cluster membership determination from \textit{Gaia} data has become a standard procedure. Different Bayesian algorithms are used to this end. These algorithms identify statistical significant groupings that exist in both spatial (coordinates) and astrometric parameter space (proper motions and parallax). There are many situations where this procedure is straightforward, but, if conditions are far from ideal, it sometimes can lead to significant confusion \citep[see, e.g. the discussion in][]{negueruela2022}. The case of NGC~663 is far from straightforward, as 1) it is embedded in an association whose members share similar astrometric parameters with the cluster; 2) it is surrounded by other open clusters with very similar astrometric parameters (see more on this in Section~\ref{sec:association}); and 3) it is a distant cluster located in a spiral arm; if proper motions are dominated by Galactic rotation, the field population of the arm may have similar astrometric parameters to cluster members. This last source of interlopers is hard to avoid, but, in the case of young open clusters, can be later corrected through analysis of the CMD. 

To address the first two issues, there are different approaches. \citet{Cantat2020} resorted to catalogue values of cluster radii and conducted searches over somewhat larger fields, as detailed in \citet[][]{canand20}. However, many of the clusters (especially the larger ones) exhibit circular halos extending to the search limit, most likely indicating that the cluster population dissolves into the association population, which has similar parameters. Conversely, {\sc hdbscan}, as implemented by \citet{hunt2023}, identifies cluster radii, and can assign members to NGC~663 and NGC~654, despite nearly identical astrometric parameters and intersecting spatial extents. The shapes reported for both clusters are, however, very asymmetric and highly irregular, suggesting again that members of the surrounding association are included. Moreover, the parameters for NGC~663 reported by both sets of authors differ almost significantly in pmDec, indicating that they are not taking exactly the same populations. A direct comparison between the two sets of members reveals that \citet{hunt2023} select many more faint ($G=16$--18) members, demonstrating the much better quality of EDR3 astrometric data with respect to EDR2, but also includes members at higher distances from the cluster centre (up to $35\arcmin$ in some directions).

Separating a cluster from its surrounding association poses significant challenges.
To mitigate them, we adopted a completely different approach, by using the following procedure. We started with the catalogue of \citet{cantat18}, who used the UPMASK code on \textit{Gaia} EDR2 data \citep{Gaia Collaboration2} within a circle of $30\arcmin$ around the nominal centre of NGC~663 to identify 1774 objects with a non-zero probability of being members of the cluster. For a better characterisation, we obtained proper motions and parallaxes in EDR3 by crossing this list with the EDR3 catalogue \citep{Gaia Collaboration3}. We then calculated median values and standard deviations for these parameters. We followed an iterative procedure to remove outliers by rejecting objects for which at least one of the three parameters (pmRA, pmDec and $\varpi$) falls outside two standard deviations from the median, and then recalculating the median values and standard deviations. The procedure converges to a list of likely members based only on parallax and proper motions, without taking into account their location in the sky. Table~\ref{tab:median} lists the median values for this population.

\begin{figure*}
\resizebox{\textwidth}{!}{\includegraphics[angle=0]{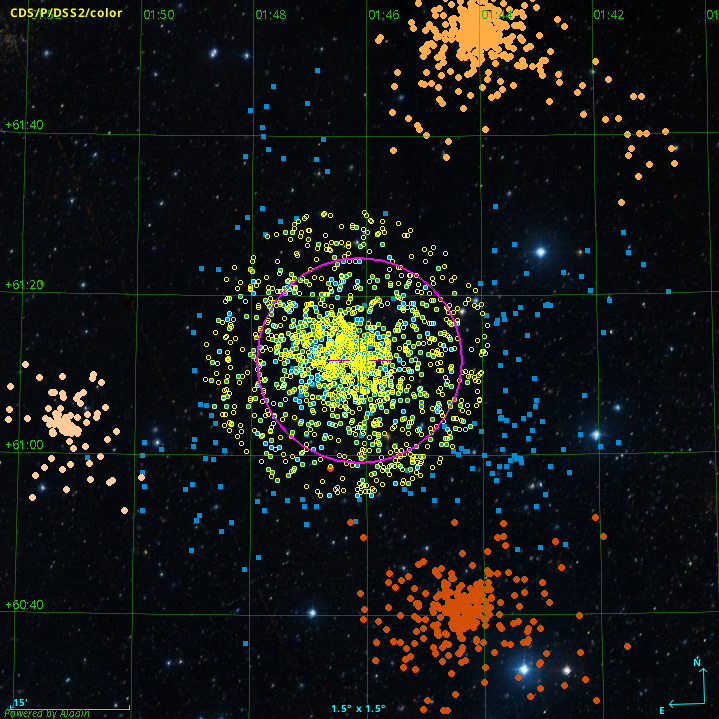}
}
\centering
\caption{Different methods to evaluate the extent of NGC~663. The yellow circles represent our selection of members (cluster + association) out to a radius of 18$\arcmin$. The filled blue squares are cluster members according to \citet{hunt2023}. Solid circles in different shades of orange are members of NGC~654 (top right), Berkeley~6 (left) and NGC~659 (bottom right) by \citet{hunt2023}. The magenta circle represents the extent of the cluster as determined with ASteCA, which may be understood as the (diffuse) limit between cluster and surrounding association. North is up and East is left. The size of the image is $1\fdg5\times1\fdg5$.\label{fig:distribution}} 
\end{figure*}

\begin{table}
	\centering
	\caption{Median values in EDR3 of parallax (uncorrected) and proper motions for the members of NGC~663.\label{tab:median}}
        \begin{tabular}{ccc}
       Plx&pmRA&pmDE\\
       (mas)&(mas$\cdot$a$^{-1}$)&(mas$\cdot$a$^{-1}$)\\
        \hline
        $0.32\pm0.10$&$-1.12\pm0.31$&$-0.26\pm0.31$\\   
        \hline
        \end{tabular}
\end{table}

We then extended the search to two concentric circles around the nominal centre of NGC~663, with radii  $18\arcmin$ and~$38\arcmin$, respectively (see Fig.~\ref{fig:distribution}). Objects in these circles were selected if their proper motions and parallaxes fell within a 2--$\sigma$ range from the medians previously defined (listed in Table~\ref{tab:median}). We find 1456 stars fulfilling these conditions within the smaller circle and 3841 stars in the outer one. This confirms that the association surrounding NGC~663 shares its astrometric parameters. Of course, the number of likely members decreases as we move out from the cluster core, but there are substantial numbers at very high distances. Specifically, the outer ring, encompassing a surface approximately 3.5 times larger than that of the inner circle, contains 1.4 times as many astrometric members.

In view of this spatial continuity, where the halo of NGC~663 encompasses NGC~654 and NGC~659, we designate the inner circle as the true extent of the cluster, i.e.\ the largest area that includes the cluster, but does not include the haloes of neighbouring clusters (see Fig.~\ref{fig:distribution}). This choice aligns well with the outcomes of more sophisticated methodologies, as we will see in Section~\ref{sec:geo}. For the 1456 objects identified as likely members, the zero point bias correction for the parallax was calculated following the procedure described by \citep{lindegren} and using their Python code\footnote{https://gitlab.com/icc-ub/public/gaiadr3\_zeropoint}. The corrected median parallax for these members in \textit{Gaia} EDR3 is $0.37\pm0.05$, which corresponds to a distance of 2.7~kpc.

 \begin{figure}
\resizebox{\columnwidth}{!}{\includegraphics[clip]{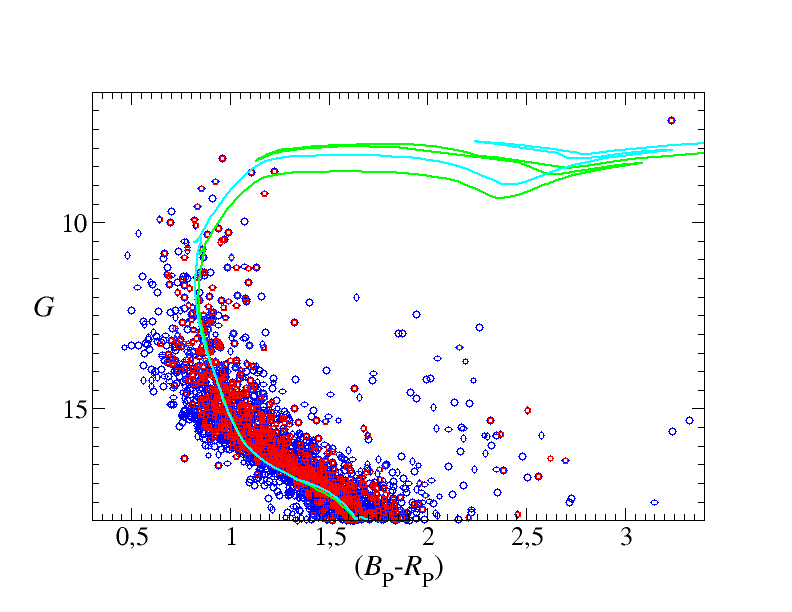}}
\caption{$G$ magnitude against ($BP-RP$) colour for \textit{Gaia} EDR3 astrometric members of NGC~663 distributed in an area of $18 \arcmin$ in radius from the nominal centre (red dots) and for an area of $38\arcmin$ in radius from the centre (blue dots). The isochrones shown are PARSEC \citep{bressan2012} isochrones for solar metallicity and ages $\log\,t= 7.3$ (blue line) and $\log\,t= 7.4$ (green line). \label{phot_Gaia}}
\end{figure}

Fig.~\ref{phot_Gaia} shows the $G$ magnitude against ($BP-RP$) colour diagram for the stars selected in both the inner circles (depicted as red dots) and the outer ring (illustrated as blue dots). Notably, the main sequences for both regions align closely, with the primary distinction being the higher incidence of objects located outside the cluster sequence within the outer circle sample. This is consistent with the anticipated rise in astrometric interlopers (i.e.\ objects in the field that happen to have similar proper motions and parallax by chance) with increasing surface, as the stellar population is essentially the same \citep[cf.][]{negueruela2023}. We confirm the validity of our member selection method by counting the number of the stars in the CMD with bad locations. For blue dots the fraction is 42/3841 and for red dots, it is 15/1456, i.e.\ a factor 2.8, directly comparable with the ratio between the areas (4.5). In both cases, $\approx1\%$ of the stars are astrometric interlopers. Therefore, assuming that some stars may fall on the cluster sequence by chance (as the reddening line follows the main sequence for \textit{Gaia} photometry), we can safely state that less than 2$\%$ of the astrometric members are contaminants.


Among the objects for which we have spectra (refer to Tables~\ref{tab:SGFIES},~\ref{tab:R4000},~\ref{tab:R1400} and~\ref{tab:R900}), we consider a star as a likely member of the cluster if its parallax and proper motions fall within the range of values shown in Table~\ref{tab:median}  during the initial selection, without applying zero point bias correction in parallax. Subsequently, all values were corrected using the method outlined by \citet{lindegren}, as described earlier. We find a small, but non-negligible, number of objects whose parallax is consistent with membership that display divergent proper motions.
The possibility that these objects are runaway members is discussed in App.~\ref{app:runaways}.

\subsection{HR diagrams}

\subsubsection{Reddening determination}
\label{sec:reddening}

Our selection of members in NGC~663 was done based on the \textit{Gaia} proper motion values (see Sect.~\ref{sec:gaia}). We have Str\"omgren photometry for 180 members covering the B-type spectral range in the $V$/$(b-y)$ and $V$/$c_{1}$ diagrams. Our main goal is to determine the physical properties of the cluster: distance and age. For this purpose, we built the $M_{V}-c_{0}$ diagram, which is known to be much more sensitive to age than the \textit{Gaia} CMD.
The individual values of $c_{0}$ were calculated using the following relations: $c_{0}=c_{1}-E(c_{1})$; $E(c_{1})=0.2\cdot E(b-y)$.
We calculated individual $E(b-y)$ for all cluster members with the procedure described by \citet{crawford1970a}, and then used the observed $c_{1}$ to predict the first approximation to $(b-y)_{0}$ with the expression $(b-y)_{0}$=$-0.116 + 0.097\cdot c_{1}$.
After this, we calculated $E(b-y)=(b-y)-(b-y)_{0}$ and used $Ec_{1}=0.2\cdot E(b-y)$ to correct $c_{1}$ for reddening $c_{0}=c_{1}-E(c_{1})$. 
The intrinsic colour $(b-y)_{0}$ was then calculated by replacing  $c_{1}$ with $c_{0}$ in the above equation for $(b-y)_{0}$. Three iterations are enough to reach convergence in the process. Restricting the analysis to stars that have never been classified as Be stars, we found a moderate degree of differential reddening amongst members, ranging from 0.52 until 0.79. The average value of the colour excess is $E(b-y)=0.61\pm0.05$ and $E(c_{1})=0.12\pm0.01$, where the error bars are given by the standard deviations of all values.

\begin{figure*}
\resizebox{\textwidth}{!}{\includegraphics[angle=0]{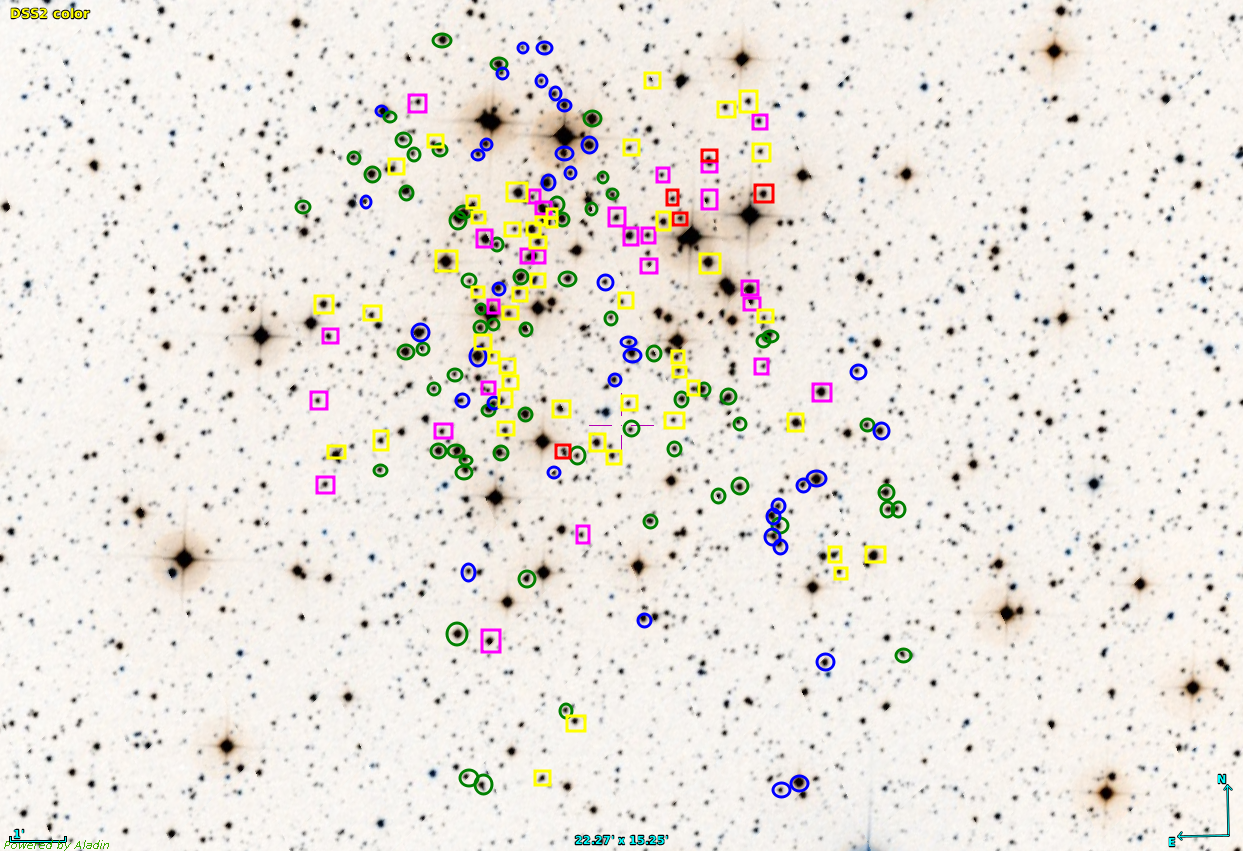}
}
\centering
\caption{Distribution of $E(b-y)$ in the core of NGC~663. Colours represent different reddening values in intervals of 0.05 magnitudes. Blue and green ellipses are the lowest values, yellow rectangles indicate values close to the average, and magenta and red rectangles are the highest values. North is up and East is left. The size of the image is $22.27\arcmin\times15.25\arcmin$\label{reddening}.} 
\end{figure*}

In Table~\ref{tab:abs} we show the values of $E(b-y)$;  $E(c_{1})$; $V_{0}$; $c_{0}$ and $M_{V}$ for all members in NGC~663. In Fig.~\ref{reddening}, we plot the different values of reddening $E(b-y)$ in the field of the cluster. Colours represent  different reddening values in intervals of 0.05 magnitudes. Blue and green ellipses are the lowest values, yellow rectangles identify values close to the average, and magenta and red rectangles are the highest values. There is a concentration of the highest $E(b-y)$ values close to the northwestern corner, while the lowest $E(b-y)$ can be found to the northeast and southwest. The central regions of the cluster are occupied by stars with mixed values of $E(b-y)$. 

\subsubsection{Determination of distance and age}
\label{sec:distance}

\begin{figure}
\resizebox{\columnwidth}{!}{\includegraphics[angle=0]{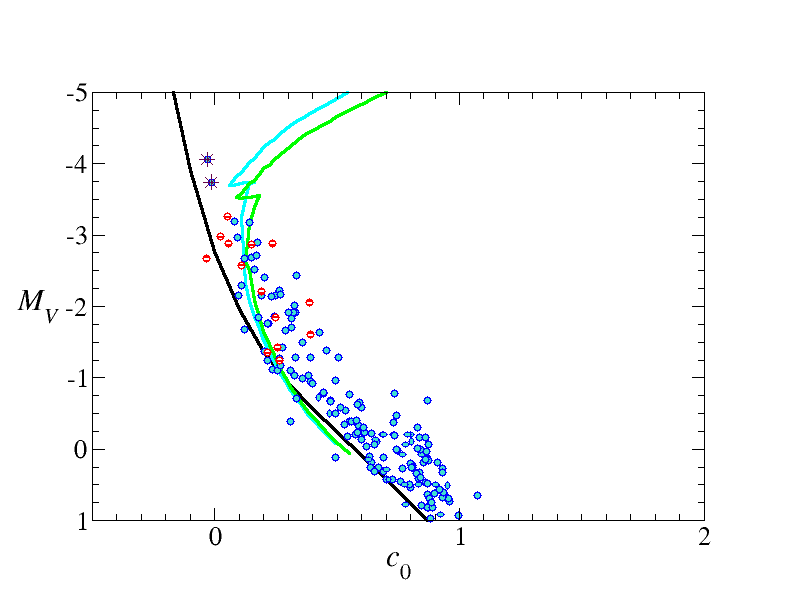}
}
\caption{Absolute magnitude $M_{V}$ against intrinsic colour $c_{0}$ for \textit{Gaia} members with Str\"{o}mgren photometry. Red striped symbols represent Be stars, whose reddening may be overcorrected. The two star symbols indicate spectroscopic blue stragglers. The black solid line is the ZAMS from \citet{Perry1987}. Two isochrones of \citep{Marigo2008}, corresponding to $\log\,t = 7.3$ (cyan) and $\log\,t = 7.4$ (green), are shown. \label{phot_spec}}
\end{figure}

Although \textit{Gaia} provides an accurate distance for NGC~663, we conducted a traditional analysis to exploit the Str\"{o}mgren data and ensure consistency.  In Fig.~\ref{phot_spec}, we illustrate the $M_{V}$/$c_{0}$ diagram for all \textit{Gaia} members with photometry from Table~\ref{tab:abs}. We estimated the distance modulus (DM) to NGC~663  by fitting the observed $V_{0}$ vs. $c_{0}$ zero age main-sequence (ZAMS) to the mean calibrations of \citet{Perry1987}. The ZAMS is fitted as a lower envelope for the members, resulting in a best-fitting distance modulus of $V_{0}-M_{V}=12.2\pm0.2$. The error represent the uncertainty in positioning the theoretical ZAMS and its identification as a lower envelope (see Fig.~\ref{phot_spec}). This DM corresponds to a distance of $2.7^{+0.4}_{-0.4}$~kpc, fully consistent with the value derived from \textit{Gaia} parallaxes (see Sect.~\ref{sec:gaia}). The photometric analysis assumes an extinction law close to the standard, validated by the consistent results for distance obtained using both independent methods. 

We determined the cluster age by visually fitting isochrones in the observational $M_{V}$/$c_{0}$ diagram at the position of the turn-off stars (see Fig.~\ref{phot_spec}). Isochrones from \citet{Marigo2008}, computed using a \citet{kroupa1998} initial mass function (IMF) corrected for binaries, were utilised. The reddening procedure uses the extinction law from \citet{cardelli1989} and \citet{odonell1994} with $R_{V} = 3.1$. The best-fitting isochrones, spanning ages between $\log\,t= 7.3$ ($\tau=20$ Ma) and $\log\,t= 7.4$ ($\tau=25$ Ma), were chosen as those that separate from the ZAMS following the turn-off stars with spectral types around B2.5\,IV, which present $M_{V}$ around $-3.1$, in agreement with calibrations \citep[e.g.][]{humphreys1984}. We note, however, that all the stars classified as giants or supergiants are saturated in our photometry, except for 30 (B1\,III), which appears as a blue straggler above and to the left of the isochrone (as also does 4, B1\,IV; cf.\ Section~\ref{sec:ums}). Other stars to the left of the isochrones at lower luminosities are Be stars, whose reddening has been overcorrected by the procedure, which does not separate circumstellar from interstellar reddening. We made sure not to include any Be star in the fit. The position of the turn-off indicates the cluster age to be $\tau=22.5 \pm 2.5$ Ma, the range permitted by the two isochrones that give a good fit to the dereddened $M_{V}$/$c_{0}$ plot.

\subsection{Geometry and size of the cluster}
\label{sec:geo}

To further probe the cluster's geometry, we used the  Automated Stellar Cluster Analysis (ASteCA) code \citep{perren2015}. We employed an input file with the EDR3 parameters of the 1456 \textit{Gaia} members determined by our procedure. Since our method does not differentiate between the cluster proper and the surrounding association with similar proper motions, ASteCA evaluates the overdensity represented by the cluster over the association.  By fitting a King profile, it obtains the following characteristics: coordinates of the cluster centre (RA=26.53$\degree$=01h 46m 6.8s; DEC=$+$61.20$\degree$=$+61\degree$~11$\arcmin$~44.1$\arcsec$) and a radius of the cluster of $12\farcm8$. Given the uncertainties, this is not significantly different from the $18\arcmin$ considered by our initial procedure, as can be seen in Fig.~\ref{fig:distribution}. Both radii are well within the area occupied by the members determined by \citet{hunt2023}.

\subsubsection{Cluster mass}
\label{sec:masscalc}

\begin{table}
\centering
\caption{Values of spectral types for B-type members on the main sequence and their mass from \citet{harmanec1988}, number of stars in the cluster, and total mass for each spectral type in NGC~663.\label{tab:mass}}
\begin{tabular}{lccc}
\hline
\hline
\noalign{\smallskip}
Spectral Type& Mass ($\mathrm{M}_{\sun}$)&Number&Total mass ($\mathrm{M}_{\sun}$)\\
\noalign{\smallskip}
\hline
\noalign{\smallskip}
B2\,V&8.6&14& 120.4\\
B2.5\,V&7.2&23&165.6\\
B3\,V&6.1&14&85.4\\
B4\,V&5.1&36&183.6\\
B5\,V&4.4&26&114.4\\
B6\,V&3.8&23&87.4\\
B7\,V&3.4&24&81.6\\
B8\,V&2.9&71&205.9\\
B9\,V&2.5&77&192.5\\
\noalign{\smallskip}
\hline
\end{tabular}
\end{table}

The calculation of the IMF with which a group of stars is formed is one of the most important issues in modern Astrophysics. For NGC~663 we can only determine the present-day mass function by counting stars.
Our selection of Gaia members extends to  $G=19$. A few more members can be found at fainter magnitudes, but completeness is unlikely. Given the values of distance modulus and average extinction that we find, stars around this limit will be slightly more massive than the Sun. We can, however, be certain that our member list is completed for the whole B-type. As discussed in \citet{negueruela2023}, the \textit{Gaia} photometric sequences of young open clusters consist of a vertical stretch corresponding to B type stars and a sloping sequence for less massive stars. The transition between these two features can be perfectly seen around $G=15.4$ in Fig.~\ref{phot_Gaia}.

We have spectral classification for 153 stars in the cluster, of which around 120 are dwarfs, extending down to B8. By using their location in the \textit{Gaia} CMD as guidance, we are able to trace the main sequence from spectral type B2\,V until spectral type B9\,V. Since we only have spectra for a subsample of the total population, we investigate the median $G$ magnitude for each  spectral type in the cluster and assign a range of $G$ magnitude for each one. Given the moderate extent of differential reddening found in Section~\ref{sec:reddening}, we can use these $G$ magnitude values to estimate a spectral type for each \textit{Gaia} member in the cluster. In Table~\ref{tab:mass}, we count the number of dwarf stars with a spectral type between B2  and B9 (we note that, in this cluster, there are no B1\,V stars). We then assign a stellar mass to each spectral type by using the calibration of \citet{harmanec1988}. 

This procedure results in a total of 308 main-sequence B-type stars. The number is fully consistent with the $\approx320$ B-type objects that can be identified if we plot the CMD for the members identified by \citet{hunt2023}.

The IMF for stars more massive than the Sun in the Solar neighbourhood was determined by \citet{salpeter1955} as:
\begin{equation}
\xi(M)=\xi_{0}M^{-2.35}
\end{equation}  
where $\xi_{0}$ is the constant which sets the local stellar density. 
In our case, taking a Salpeter law, we can determine the number of stars ($N$) with masses between $M_{1}$ and $M_{2}$, by integrating the IMF between these limits: 

\begin{equation}
N=\int_{M_{1}}^{M_{2}} \xi(M) dM = \xi_{0}\int_{M_{1}}^{M_{2}} M^{-2.35} dM \\ 
=\frac{\xi_{0}}{1.35}[M_{1}^{-1.35} - M_{2}^{-1.35}]
\label{eq:number}
\end{equation}

To calculate the value of $\xi_{0}$, we use the results of Table~\ref{tab:mass} to count 308 stars between~2.5 and~$8.6\:$M$_{\sun}$, the range where we are certain of completeness. We obtain a value of $\xi_{0}\approx1766$. To test the robustness of this value, we repeat the calculation for stellar mass (i.e. calculating $\xi_{0}$ by normalising the total mass between $2.5\:\mathrm{M}_{\sun}$ and  $8.6\:\mathrm{M}_{\sun}$ to $1236.8\:\mathrm{M}_{\sun}$), and for different spectral-type subsets taken from Table~\ref{tab:mass}. In all cases, the value of $\xi_{0}$ is consistent within 10\%, giving us confidence on both our completeness and the spectral-type to mass calibration used. 

Now, we estimate the total current mass of the cluster by using this value of $\xi_{0}$. The total mass in stars born with mass $M_{1}=8.6\:$M$_{\sun}\geq M \geq M_{2} =1\:$M$_{\sun}$ should be:

\begin{equation}
M=\int_{M_{1}}^{M_{2}} M \xi(M) dM \\
=\xi_{0}\int_{1 \mathrm{M}_{\sun}}^{8.6 \mathrm{M}_{\sun}} M^{-1.35} dM = 2\,670\:\mathrm{M}_{\sun}\\
\label{eq:mass}
\end{equation} 

This result gives the present-day mass of the cluster down to solar-type stars. To estimate the low-mass component, we can use a \citet{kroupa2003} standard IMF, for which the mass in low-mass stars is comparable to the mass in stars more massive than the Sun. This means that NGC~663 is a massive cluster with a total current mass close to $5\,300\:\mathrm{M}_{\sun}$.

 The effect of unresolved binarity on this number is difficult to assess. The rough approximation used by \citet{tadross2005} of assuming a binary frequency of 50\% (which is a sensible assumption for a spectroscopically confirmed sample of B-type stars) and a typical mass ratio of 0.77 implies increasing the cluster mass by 38\%. This would mean a total current mass close to $7\,000\:\mathrm{M}_{\sun}$. 

The ASteCA code provides an independent estimation of the cluster's total mass that does not rely on spectral types, but the luminosity function and a binary model. Since ASteCA was run with a pre-selected sample of association members, its mass only measures the overdensity of the cluster over the association. The code provides a total mass of $5\,340\:\mathrm{M}_{\sun}$ for a $\log\, t$=7.396$\pm$0.039 and a distance modulus of 12.19$\pm$0.08, all values consistent with our previous determination. The somewhat lower mass is due to the removal of a baseline corresponding to the density of the surrounding association. 

Nevertheless, these estimates are, in all likelihood, underestimations, as the high-mass population of such a massive cluster cannot be ignored. Extrapolation of Eq.~\ref{eq:number} from $8.6\:\mathrm{M}_{\sun}$ to $120\:\mathrm{M}_{\sun}$ indicates that the cluster should have about 70 stars evolved away from the main sequence. Given that we find 15 giants and 7 supergiants, about 50 stars should already have been lost to supernova explosions. Of this, about 25 should have been more massive than $18\:\mathrm{M}_{\sun}$ and thus started their lives as O-type stars. Integration of Eq.~\ref{eq:mass} up to $120\:\mathrm{M}_{\sun}$ adds around $1\,500\:\mathrm{M}_{\sun}$ to the 2\,670 previously calculated. Accounting for binarity and repeating all the calculations, the initial mass of the cluster is likely to have been at least $11\,000\:\mathrm{M}_{\sun}$ -- ignoring dynamical ejections.




\subsection{The Be star content}
\label{sec:becontent}

The high Be content of NGC~663 has been remarked by a number of authors, although this may simply be a consequence of the very high number of B-type stars \citep{yu2015}. Many members have been catalogued as Be stars in the literature, but \citet{sanduleak1990} warned about potential confusions and provided a clean list of likely Be stars. More recently, \citet{yu2015} provided an updated list. We have spectra of 28 of the 31 objects in their table~2, which lists Be stars from literature (we are missing star 14 in the cluster centre, and 207 in the halo, as well as 417, which is not an astrometric member). Of the 28 objects with spectra, one is a foreground G-type star (PKK1 = 240) and one is not an astrometric member (288), although ejection from the cluster is not impossible. Two bright catalogued Be stars do not show emission lines in our spectra, and there are reasons to believe that they never have, while we find 3 certain and 2 likely new Be stars. A discussion of individual cases is presented in App.~\ref{app:bes}.

In total, our spectra identify 27 Be star members, with 2 further likely Be stars. There are two other member Be stars in \citet{yu2015} for which we do not have spectra and 3 astrometric non-member Be stars that could be connected to the cluster. It is difficult to translate these detections into a Be fraction, because our list of spectroscopic targets, produced long before \textit{Gaia}, was much more complete for Be stars than for non-emission objects. For stars of luminosity class IV and III, leaving aside blue stragglers, there are 9 Be stars among approximately 30 objects. Among B2\,--\,B2.5 main sequence stars, there are eight Be stars out of about 40 objects. Among B3\,--\,4 main sequence stars, there are seven Be stars out of about 50 objects. There are three (perhaps four) B5--6\,V Be stars (out of about 50) and one or two of later spectral type (more than 150 stars). Therefore, the Be fraction in NGC~663 is not particularly high, only becoming close to 30\% above the turn-off and quickly falling off at lower masses. Other clusters have Be fractions approaching 40\% around the turn-off \citep[e.g.][and references]{hastings2021}.

\subsection{Gaia results for CasOB~8}
\label{sec:association}

Beyond its intrinsic interest, NGC~663 is the largest cluster in the association Cas~OB8. The existence of this association has been recognised for a long time.  \citet{McCuskey64} studied five major open clusters in the region (Trumpler~1, NGC~581, NGC~654, NGC~659 and NGC~663) and, despite estimating distances between 1.8\,--\,2.3~kpc, concluded that, given the uncertainties, they "... are the same distance from the sun and are in the Perseus arm of the galaxy". \citet{humphreys1970} described Cas~OB8 as "a large, distant association ... It appears centered on the cluster NGC~663". Later catalogues \citep[e.g.][]{humphreys1978, garmany1992} placed it in the Perseus Arm, around Galactic coordinates $l=129.5\degree$, $b=-00.9\degree$. 

\begin{figure*}
\resizebox{0.9\textwidth}{!}{\includegraphics[angle=0]{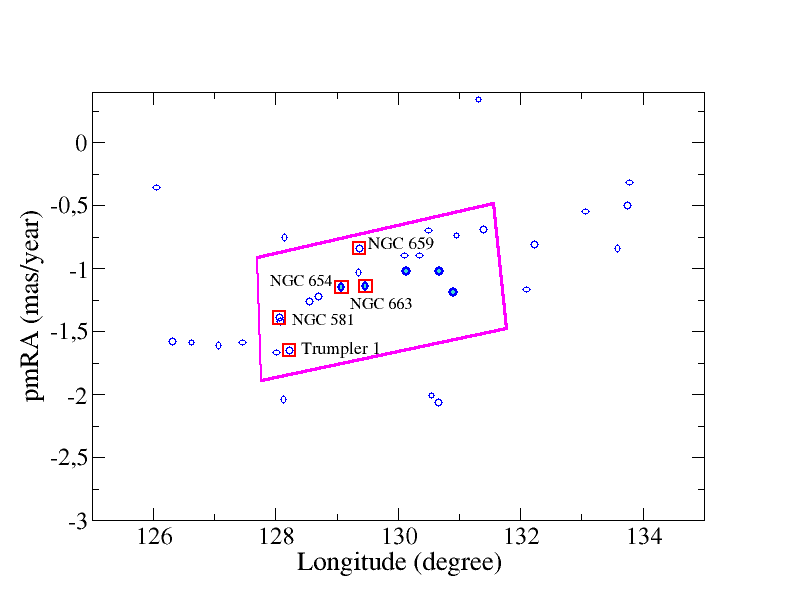}
}
\centering
\caption{Distribution of open clusters distributed between $l=134\degree$ and $l=126\degree$ around the region of Cas~OB8 ($l=129.5\degree$ and $b=-00.9\degree$. The classical open clusters belonging to Cas~OB8 are marked with red open squares and they are labelled with their names. The cyan solid circles represent to clusters with values of proper motions in RA and DEC compatible with NGC~663. Inside the magenta line are the clusters belonging to the association Cas~OB8 taken into account the values of EDR3 from \citet{hunt2023}\label{CASOB8}.} 
\end{figure*}    

Modern \textit{Gaia} data provide an excellent opportunity to reassess the physical connection between clusters in CasOB~8 and determine its distance and extent. To this aim, we select open clusters from the catalogue of \citet{hunt2023} within $126\degree <l<134\degree$ and $-7.0\degree<b<7.0\degree$, a much broader area than in previous studies. For example, \citet{humphreys1978} considered a more restricted range [$129.2\degree<l<129.7\degree$,  $-1.5\degree<b<-0.2\degree$]. This region contains 154 clusters, but a very rough initial cut to $\varpi \leq 0.5$ (to place them at least at the distance of the Perseus arm) and $\log\, t >8$ (to make sure that they are young enough to belong to an OB association) reduces the sample to 52 clusters. 

NGC~663, the largest confirmed member of CasOB~8, has $\varpi = 0.35\pm0.03$ and proper motions pmRA = $-1.14\pm0.08$ and pmDec = $-0.34\pm0.09$~mas/a, respectively \citep {hunt2023}.  we selected clusters with parallax values consistent (within uncertainties) with NGC 663, yielding 34 candidates (listed in Table~\ref{tab:clusters_cas0b8}) along with their Galactic coordinates, parallaxes, proper motions, and ages from \citet{hunt2023}.   Fig.~\ref{CASOB8} plots all these clusters in a Galactic longitude vs. pmRA diagram, marking classical Cas~OB8 members in red and additional clusters with cyan circles.  The magenta polygon delineates a very compact and well defined structure, which is also evident in the vector point diagram. There are 18 clusters within this polygon, which we identify as the members of Cas~OB8. Their names are labelled in bold in Table~\ref{tab:clusters_cas0b8}.

\section{Discussion}
\label{discussion}

\subsection{Cluster mass}
\label{disc:mass}

Based on two different methods, in Section~\ref{sec:masscalc}, we estimate a current mass for NGC~663 well in excess of $5\,000\:\mathrm{M}_{\sun}$, with an implied initial mass likely in excess of $10\,000\:\mathrm{M}_{\sun}$. As further confirmation, \citet{hunt2024} estimate a current mass of $7\,800\pm 800\:\mathrm{M}_{\sun}$ for NGC~663, in excellent agreement with our values. This cluster mass is much higher than previously assumed, and places NGC~663 slightly above the threshold for massive young clusters, as defined by \citet{portegies2010}. Traditionally, NGC~869 ($h$~Per) has been identified as the most massive cluster in the Perseus arm. In fact, \citet{portegies2010} gives mass estimates for both NGC~869 and NGC~884 (the $h$ and $\chi$~Per double cluster) around $10^{4}\:\mathrm{M}_{\sun}$. This value, however, is dependent on the area surveyed, as both clusters lie close to the centre of the rich Per~OB1 association \citep{deBurgos2020}, just like NGC~663 is within Cas~OB8. By looking only at the clusters themselves, \citet{slesnick2002} estimated masses around $3\,700$ and $2\,800\:\mathrm{M}_{\sun}$, respectively, for stars more massive than the Sun. These values are considerably reduced by the recent estimate, based on \textit{Gaia} EDR3 membership, by \citet{hunt2024}, who assign only $2\,400\pm 400$ and $1\,650\pm 330\:\mathrm{M}_{\sun}$ for $h$ and $\chi$~Per, respectively.

The validity of these numbers can corroborated by simply plotting the \textit{Gaia} CMD for all the cluster members selected by \citet{hunt2024} and identifying B-type stars by selecting the vertical tract of the sequence, as detailed by \citet{negueruela2023}. The number of B-type stars in $h$~Per is around 130, while in $\chi$~Per, it is around 90. This must be compared to the approximately 320 B-type stars found in NGC~663, even though it is an older cluster. NGC~663 is thus more massive than the whole central concentration of Per~OB1. The only cluster in the Perseus arm with a comparable mass is NGC~7419, which contains over 200 B-type stars \citep{marco2013,hunt2024} The very high number of BSGs and Be stars in NGC~663 can be simply explained as a direct consequence of its mass. Because of this huge population of moderately massive stars, NGC~663 is an ideal laboratory to test models and ideas about the evolution of such stars, as the samples may be statistically significant.

\subsection{Upper main sequence population}
\label{sec:ums}

\begin{figure*}
\resizebox{\columnwidth}{!}{\includegraphics[angle=0]{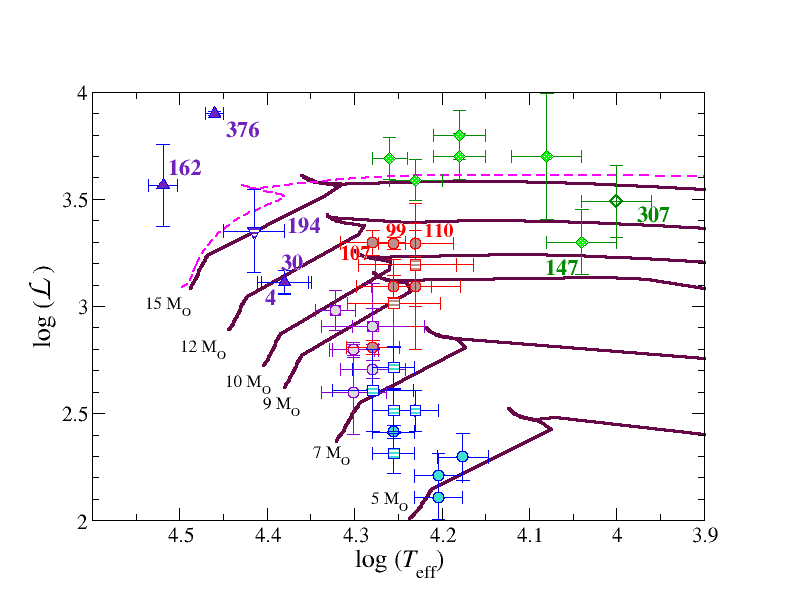}}
\resizebox{\columnwidth}{!}{\includegraphics[angle=0]{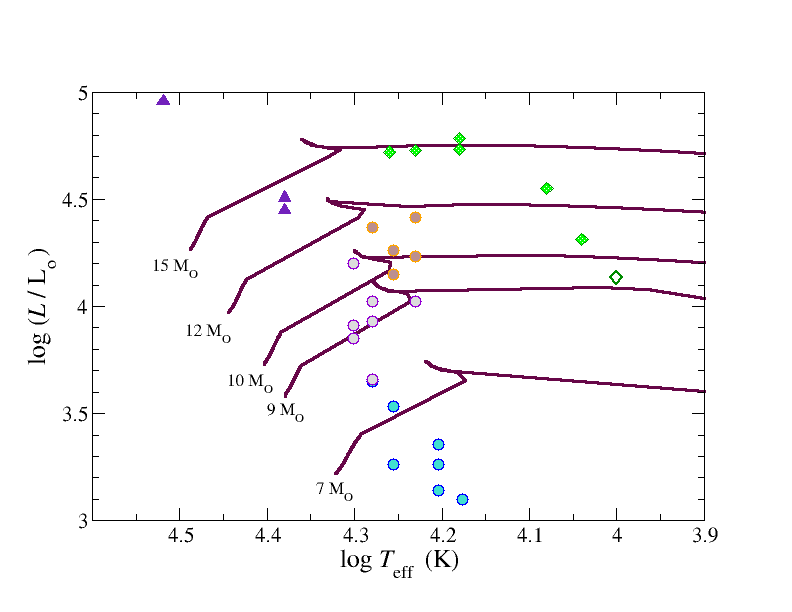}}
\caption{\textbf {Left:} Spectroscopic HR diagram for stars in NGC~663. Stars classified as luminosity class V are shown in turquoise, luminosity class IV in purple and giants in brown (in all cases, Be stars as striped squares and non-emission stars as filled circles). The spectroscopic blue stragglers are shown as triangles (the stripped triangle is the Be star 194). Objects classified as supergiants are shown as green diamonds. The continuous lines are evolutionary tracks with standard rotation from \citet{brott2011}, labelled with the stellar mass. The dashed purple line is the standard-rotation $15\:\mathrm{M}_{\sun}$ track from \citet{ekstrom2012}, shown for comparison. \textbf{Right: } Standard HR diagram for the subset of objects that have reliable $UBV$ photometry in the literature -- with Be stars excluded, because of their variability. The luminosity has been calculated by assuming our distance to the cluster, a standard extinction law and the bolometric correction corresponding to the best-fitting model. Note that 376 (BD~$+60\degree$\,345), which is an astrometric member in the halo, lies outside the plot at $\log\,\left(L/L_{\sun}\right)\approx 5.4$. In both diagrams, the open diamond represents the position of 307 (BD~$+60\degree$\,336), which is likely an ejection from the cluster. \label{fig:brottdiags} }
\end{figure*}     

The cluster turn-off is sharply defined in our spectroscopic sample, with no stars classified as dwarfs earlier than B2. When total numbers of stars with spectra in Tables~\ref{tab:R4000}, \ref{tab:R1400} and~\ref{tab:R900} are considered, we count about 15 stars with spectral type B2 and a luminosity class ranging from V to III, alongside about 25 classifed B2.5, covering the same luminosity range. Finally, there is a single B3\,III giant. As seen in Sect.~\ref{sec:masscalc}, when stars for which only \textit{Gaia} photometry is available are included, using their brightness as a proxy, we arrive at a total of 14 B2\,V stars and 23 B2.5\,V stars (Table~\ref{tab:mass}). This distribution of spectral types aligns very well with the photometrically determined age of 23~Ma at solar metallicity (Sect.~\ref{sec:distance}) with the turn-off indicated by the substantial population of B2.5\,IV stars. A directly comparable distribution is observed in the massive SMC cluster NGC~330 \citep{bodensteiner2020}, although it may peak at slightly later types due to limitations in their data (they cannot separate B2.5 or luminosity class IV). The rather older age inferred for NGC~330 (35\,--\,40~Ma) is mostly attributable to the much lower metallicity. 

In NGC~663, the small spread in spectral types for stars around and above the turn-off can be readily explained by different initial rotation velocities. A distribution of rotation velocities leads to the extended turn-offs seen in most massive young open clusters, as stars that rotate faster have longer main-sequence lifetimes and appear to be younger than non-rotating stars of the same age \citep[e.g.][]{niederhofer2015}. Be stars further lead to the spread in the CMD, as they tend to have redder colours than stars without emission lines. On the other hand, all stars earlier than B2 can be considered spectroscopic blue stragglers (SBSs) in NGC~663. Interestingly, none of them is given a luminosity class V. The five SBSs are:
\begin{itemize}
    \item The earliest SBS is 162 (BD~$+60\degree$\,329), O9.2\,IV, which is in the halo to the northwest. This star is only $\approx 8\arcmin$ from the centre, but its kinematics suggest that it is a member of the association, rather than the cluster. Nevertheless, our low-resolution spectrum indicates that this object is a fast rotator, which strongly suggests that it is the product of binary interaction \citep[cf.][]{holgado22}.
    \item To the north, at the edge of our $18\arcmin$ circle, lies 376 (BD~$+60\degree$\,345), which has astrometric parameters fully consistent with membership. At $G=9.6$, this B0\,III star is only half a magnitude fainter than the faintest cluster supergiants (cf.\ Table~\ref{tab:SGFIES}). When the bolometric correction corresponding to its much higher temperature is taken into account, this is by far the most luminous star in the area (see Fig.~\ref{fig:brottdiags}).
    \item Near the cluster centre lie two mild SBSs: 4 (B1\,IV), a $\beta$~Cep pulsator (V1155~Cas), and 30 (B1\,III, although it is close to the lower luminosity boundary for the class, only 0.3~mag brighter than 4). These are the two stars clearly above and to the left of the turn-off in Fig.~\ref{phot_spec}, as they are the only SBSs within the photometric field. 
    \item Finally, the most interesting SBS is 194 (LS~I~$+61\degree$\,235), a B0.7\,IVe shell star and the optical counterpart of the X-ray binary RX~J0146.9+6121, which we will discuss below. 
\end{itemize}

In addition to this population, six BSGs have astrometric parameters compatible with membership (Table~\ref{tab:SGFIES}). Four are close to the cluster centre, while two are found in the halo to the East. They cover the range B2.5 to B9 at luminosity class Ib. A seventh BSG (BD~$+60\degree$\,336; A0\,Ib) lies to the North of the cluster and has astrometric data compatible with an ejection from the cluster. One red supergiant, BD~$+60\degree$\,335 (V589~Cas; M3\,Iab-Ib) is a halo member to the South. A second red supergiant in the same area, BD~$+60\degree$\,327 (M0\,Iab) has astrometric parameters compatible with an ejection from the cluster core. This object is likely to be a very long period binary \citep{mermilliod2008}. This is one of the largest complements of evolved stars associated with a Galactic cluster.

To understand the relationship between all these objects, we resort to the spectroscopic HR diagram \citep[sHRD;][]{langerku2014,castro2014}, which solely relies on spectroscopic parameters and is thus free of complications due to variable reddening. In Fig.~\ref{fig:brottdiags}, the left panel illustrates the positions of all stars for which we have obtained stellar parameters, with the exception of the three Be stars with unexpectedly high $T_{\mathrm{eff}}$. The turn-off stars (which we identify with spectral types B2\,V and B2.5\,IV) have masses between $8$ and $9\:\mathrm{M}_{\sun}$, while the giants extend between 9~and $11\:\mathrm{M}_{\sun}$, which is in general agreement with the calibration used in Table~\ref{tab:mass}. SBSs also align closely with tracks suitable for their spectral type. In particular, 162 is compatible with a mass close to $20\:\mathrm{M}_{\sun}$, a value considerably higher than that of any other cluster member but within expectations for its spectral type. Interestingly, only one of the six BSGs falls close to the same track as the giants, namely BD~$+60\degree$\,347, which is significantly less luminous than the rest and lies between the 10~and $12\:\mathrm{M}_{\sun}$ tracks. All other supergiants are compatible with a mass somewhat above $15\:\mathrm{M}_{\sun}$.

The high masses inferred for these stars pose a challenge within the context of single evolutionary models. For the estimated age of the cluster, stars more massive than $\approx11\:\mathrm{M}_{\sun}$ should have already undergone supernova, according to \citet{Marigo2008} isochrones. Even with very high rotation Geneva \citep{ekstrom2012} isochrones, stars more massive than $\approx12.5\:\mathrm{M}_{\sun}$ are expected to have exploded by $t=20\:$Ma, the lower limit permitted by isochrone fit and turn-off spectral type. This circumstance, however, is not unique to NGC~663, but is prevalent among Milky Way clusters of similar ages \citep[cf.][]{negueruela2017}. 

The situation differs from that found in the analogous SMC cluster NGC~330 \citep{bodensteiner2020} in two key aspects. Firstly, all the BSGs in NGC~330 are  of A-type, significantly cooler than those in NGC~663. Secondly, at the lower metallicity of the SMC, evolutionary tracks for stars of $8$\,--\,$12\:\mathrm{M}_{\sun}$ exhibit pronounced blue loops extending to the position of A-type supergiants. In contrast, at Milky Way metallicity, blue loops only reach the position of F-type supergiants and disappear entirely for masses $\ga11\:\mathrm{M}_{\sun}$ (\citealt{anderson2014}; see also \citealt{negueruela2020}). This implies that all the BSGs in NGC~330 can naturally be explained as the product of blue loops, whereas none of the BSGs in NGC~663 (or other Galactic clusters) can. Given that all these high-mass objects are confirmed astrometric members of the cluster, explaining their presence necessitates a mechanism that does not involve significantly altering their space velocity. 

\begin{figure}
\resizebox{\columnwidth}{!}{\includegraphics[angle=0]{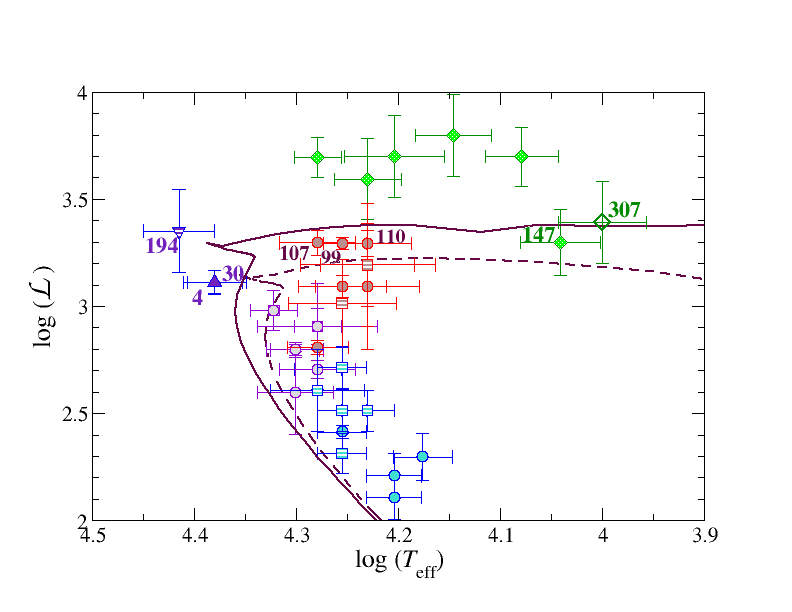}
}
\caption{Spectroscopic HR diagram for stars in NGC~663, as in Fig.~\ref{fig:brottdiags} (right panel), with isochrones from the thick-grid models of \citet{georgy2013}. The continuous line is the $\log\,t=7.3$ isochrone with initial $\Omega/\Omega_{\mathrm{crit}}=0.6$. The broken line is the corresponding $\log\,t=7.4$ isochrone.\label{fig:shrd_isos}}
\end{figure} 

The reliability of the masses derived from the location of stars in the sHRD is a crucial point that requires careful evaluation.  
As a first test, we construct the HR diagram for those cluster members with reliable literature photometric measurements in the $UBV$ sourced from the WEBDA database (Fig.~\ref{fig:brottdiags}; right panel). We assume a standard extinction law with $R=3.1$ and our distance to the cluster. Disagreement between these two diagrams is generally interpreted as an indication of anomalous stellar evolution. However, the masses inferred from the sHRD are fully compatible with those derived from the HR diagram for nearly all stars. The only exception is the cooler supergiants, which have lower masses in the HR diagram, very likely because of incorrect bolometric corrections, as their temperatures cannot be considered acccurate, lying very close to the edge of our grid.

In addition, the masses derived from both spectroscopic diagrams are also in excellent agreement with those obtained from the isochrones used to fit the Str\"{o}mgren photometry. As further confirmation, in Fig.~\ref{fig:shrd_isos}, we present once again the cluster sHRD, this time overlaid with the Geneva isochrones for solar metallicity and initial $\Omega/\Omega_{\mathrm{crit}}=0.6$ \citep{georgy2013} overplotted. As before, in this plot stars above the turn-off fall between the $\log\,t=7.3$ and~$7.4\:\mathrm{Ma}$ tracks, while the supergiants are positioned well above any of them. Nevertheless, the location of the turn-off stars agrees much better with the extended main sequence of the \citet{brott2011} tracks.

The strong agreement across independent datasets (Str\"{o}mgren photometry and spectroscopic parameters) and stellar models (Geneva vs.\ Padova) is remarkable, giving full credibility to the stellar masses obtained. Given this consistency, a mechanism must be proposed to explain the offset position of the majority of BSGs in the various diagrams. It may be argued that their positions in the diagrams do not imply the masses that are derived from standard single-star evolution tracks. For example, \citet{renzo2023} conclude that rejuvenated accretors have less bound envelopes and are subject to pronounced blue loops that can take them to the region of BSGs -- although their models assume very low ($\mathrm{Z}_{\sun}/10$) metallicity. Nevertheless, even if the BSGs are  indeed overluminous and have smaller masses than implied by their position on the tracks, this scenario still necessitates mass transfer in a binary. 

Before exploring the different evolutionary paths that may lead to this population, we will first examine the one blue straggler for which we have a clear indication of the process that led to its formation.

\subsubsection{The Be/X-ray binary}
\label{sec:bex}

The X-ray source RX~J0146.9+6121 is an accreting neutron star orbiting  LS~I~$+61\degree$\,235 (star 194 in NGC~663). X-ray pulsations reveal a  spin period of $\approx25$~min, possibly the longest period of any X-ray pulsar in a Be-star system \citep{Mereghetti2000}. RX~J0146.9+6121 is a member of a rare subclass of Be/X-ray binaries known as persistent, low-luminosity Be/X-ray binaries \citep[see][for a review of Be/X-ray binaries and their subtypes]{reig2011}. As suggested by its name, this subclass is characterised by persistent X-ray luminosity at a low level (typically, $L_{\mathrm{X}}\approx10^{34}$\,--\,$10^{35}\:\mathrm{erg}\,\mathrm{s}^{-1}$), as opposed to the bright outbursts that characterise most Be/X-ray binaries. Other characteristics of the type include long pulse periods, very weak 6.4~keV iron lines in the X-ray spectra and possibly a soft thermal excess in the X-ray spectrum. This set of characteristics may be interpreted in terms of systems with wide orbits presenting low eccentricities, again in opposition to standard outbursting Be/X-ray binaries, which generally have moderate or large eccentricities. The orbital period of RX~J0146.9+6121 is not known, but is expected to be long, as the lengths of spin periods and orbital periods are generally correlated in Be/X-ray binaries.

Located in the halo of NGC~663, star 194 is a mild SBS. \citet{reig2000} classified it as B1\,V and here we refine the classification to B0.7\,IV, which is consistent with the mass of $\approx14\:\mathrm{M}_{\sun}$ derived from its position in the sHRD (although we note that the errors are large, because of its Be nature). \citet{reig2000} detected large changes in the strength and shape of the emission lines, with a period close to 1240~d. Based on different periodicities found in its lightcurve, \citet{sarty2009} suggested that it could be a $\beta$~Cep pulsator and that the orbital period might be close to 330~d.

With a parallax exactly coincident with the cluster average, star 194 is a cluster member according to our criteria, although it is not selected as such by \citet{hunt2023}. If we subtract the cluster average proper motions from those of star 194, the residuals (though smaller than the standard deviation for cluster members) are consistent with a very slow ejection from the cluster centre. The existence of a neutron star in the binary implies a supernova explosion at some point in the evolution of the system. The presumed low eccentricity and very small peculiar velocity\footnote{In a recent analysis, \citet{zhao2023} calculate a very low peculiar velocity of $12^{+8}_{-4}\: \mathrm{km}\,\mathrm{s}^{-1}$ relative to local Galactic rotation for RX~J0146.9+6121. However, the relevant quantity here is the peculiar velocity with respect to the parent cluster \citep[cf.][]{renzo2019}, which is $\la5\: \mathrm{km}\,\mathrm{s}^{-1}$.} suggest that the kick received during the explosion was very small. \citet{pfahl2002} proposed that low eccentricity Be/X-ray binaries could be explained by a formation mechanism involving estable mass transfer from the progenitor of the neutron star while it still keeps its radiative envolope (which happens in early Case B and early Case C mass transfer). As a consequence of this process, the original secondary, which will still be a main sequence star, will be rejuvenated (i.e.\ become a blue straggler), while the core of the mass donor will become exposed and acquire a fast rotation that will prevent a significant kick.

Of course, there are many different avenues that can lead to a particular system having low eccentricity \citep[see recent discussion in][]{larsen2024}. Nevertheless, for persistent Be/X-ray binaries, the properties of the whole subclass require a wide orbit with low eccentricity, making a supernova explosion with a weak kick necessary \footnote{In apparent contradiction, \citet{knigge2011} indicate a possible tendency of long period Be/X-ray binaries to have high eccentricities. There is, however, a very obvious observational bias in their data, as only systems with outbursts have orbital solutions and thus measurements of eccentricity, except for X~Persei, and long period systems with low eccentricity do not display outbursts.}. Possible scenarios leading to this weak kick were explored by \citet{pods2004}, who argued that initial primaries with masses in the $8$\,--\,$11\:\mathrm{M}_{\sun}$ range may experience electron-capture supernovae, which will lead to small kick velocities \citep[but see also][who find that much higher masses are necessary]{poelarends2017}. A second possibility suggested by \citet{pods2004} is the collapse of a small iron core that leads to a fast supernova explosion. Systems with such wide, low-eccentricity orbits may represent viable progenitors to double neutron star binaries \citep{tauris2017}.

The properties of LS~I~$+61\degree$\,235 can shed light on this evolutionary channel. NGC~663 presents a current turn-off mass somewhat above $8\:\mathrm{M}_{\sun}$, while stars with masses somewhat above $10\:\mathrm{M}_{\sun}$ are evolving towards the RSG phase. If we assume that the binary was ejected from the cluster centre, the flighttime to its current location is between 2 and 3~Ma. Since evolutionary models suggest that a system such as  RX~J0146.9+6121 is likely to have formed via stable mass transfer when the initial primary was evolving away from the main sequence, but still had a radiative envelope (\citealt{pfahl2002}; and see also the exploration of parameter space in \citealt{vinciguerra2020}, although this is done at SMC metallicity), the progenitor of the neutron star was very likely a star of 10\,--\,$11\:\mathrm{M}_{\sun}$ (depending on its initial rotational velocity) while the progenitor of LS~I~$+61\degree$\,235 was rather less massive, so that (semi)conservative Case ~B mass transfer may have given it its current mass of 12\,--\,$14\:\mathrm{M}_{\sun}$. Just before the supernova explosion, the system must have resembled a somewhat more massive version of $\varphi$~Per, a binary system, likely a member of Melotte~20, containing a $9.6\:\mathrm{M}_{\sun}$ $\sim$B1\,V Be star and a $1.2\:\mathrm{M}_{\sun}$ hot subdwarf \citep{mourard2015}. \citet{schootemeijer2018} have explored tailored models to explain $\varphi$~Per, concluding that mass transfer is likely to have been close to conservative. If we apply this scenario to RX~J0146.9+6121, the original primary, with an initial mass below $11\:\mathrm{M}_{\sun}$, exploded as a supernova after extensive mass loss, while the original secondary, whose initial mass was well below the supernova limit, has gained enough mass to undergo a second explosion in the future.

\subsubsection{The binary path to Be stars}
\label{binbes}

As we see in NGC~663, Be stars in open clusters are found with very high preference around the turn-off and just below it, with little evidence of a significant fraction of Be stars among stars close to the ZAMS \citep{mcswain2005}, leading to the conclusion that the Be phenomenon is somehow related to evolution. Although solvent models to explain this effect from the rotational evolution of isolated stars exist \citep[e.g.][]{granada2013}, there are strong reasons to believe that binary interaction may play a role in the formation of Be stars. After all, as we have seen in the previous section, there is an effective process to create Be stars through mass transfer that can explain Be/X-ray binaries and Be + sdO binaries \citep[][and references therein]{pols1991}. In addition to Occam's razor, the extreme rarity of binaries containing a Be star and a main sequence normal B star \citep{bodens2020Be} argues for the binary channel representing the primary, or even only, mechanism to produce Be stars. Furthermore, measured surface nitrogen abundances in many Be stars seem incompatible with the evolution of a fast rotator. For instance, \citet{Dufton2024} analysed the nitrogen abundances of Be stars in 30~Doradus and compared them to models, finding that observations were inconsistent with all the targets having a single star evolutionary history. At least 30\% of the sample was suggested to have formed via binary interaction. 

\citet{wang2023} propose that a fraction of the initially faster-rotating stars may be able to reach near-critical rotation at the end of their main-sequence evolution and produce Be stars in isolation around the turn-off. Contrarily, based on the results of \citet{granada2013} and \citet{hastings2020}, they argue that the substantial numbers of Be stars found below the turn-off advocates for a crucial role of binary interaction in creating Be stars. In their view, slow pre-mass-transfer rotation and inefficient accretion allows for mild or no enrichment even in critically rotating accretion-induced Be stars. Nevertheless, throughout the years, most population synthesis model experiments aimed at testing the contribution of the binary channel have consistently found that not all Be stars can be formed in this way \citep[e.g.][]{pols1991,vanBever1997}. More recently, \citet{hastings2021} made use of a number of assumptions to claim that up to 30\% of B-type stars may become Be stars through the binary channel.

The properties of Be stars in and around NGC~663 do not seem consistent with the same evolutionary scenario used to explain RX~J0146.9+6121. Firstly, except for LS~I~$+61\degree$\,235, no Be star is a SBS. The distribution of spectral types among the Be stars coincides with that of non-emission stars. No other SBS presents emission lines. Secondly, if case B is dominant for mass transfer, given the current turn-off mass of the cluster and the example of LS~I~$+61\degree$\,235, we would expect a significant fraction of the Be stars formed via mass transfer in a binary, which would all have primary masses $\ga9$, and most likely, $\ga11\:\mathrm{M}_{\sun}$, to have experienced the supernova explosion of the original primary and thus received a kick. Contrarily, almost all the Be stars have proper motions consistent with cluster membership and show no evidence of a perturbed velocity. Apart from LS~I~$+61\degree$\,235, the only Be star that might have been ejected from the cluster is the B3\,IVshell star 288, and only if it originates from the Eastern halo. This is not a SBS, neither particularly bright. 

It might be argued that the slow ejection of LS~I~$+61\degree$\,235 is an exception and that supernova kicks eject Be stars at higher velocities. However, we have searched the lists of emission-line stars out to 2 degrees away from the cluster centre without finding any strong candidate for an ejected star. BD~$+60\degree$\,307, about $34\arcmin$ to the West of the cluster centre has astrometric parameters typical of cluster members. If it is an ejection, it is extremely slow. The absence of candidates only allows for stars with very high peculiar velocities which are now outside our search radius. Therefore the properties of Be stars in and around NGC~663 suggest that the vast majority were formed without a supernova explosion.

\subsubsection{How are blue stragglers and BSGs formed?}
\label{sec:straggle}

To summarise, NGC~663 hosts a large population of Be stars (although not a high Be fraction), most of which show kinematics typical of cluster members, a Be/X-ray binary walkaway, three\footnote{Given that the most extreme SBS, star 162, is unlikely to be a cluster member, but rather part of the surrounding association, we will not discuss it.} non-emission SBSs (one in the halo and two in the centre) also with typical cluster kinematics, and six BSGs (plus one likely ejected) that, with one exception, are much brighter and massive than any other member. Our previous discussion has explored potential evolutionary pathways for the emission-line stars. If we want to explain the stars that look decidedly earlier and more massive than the turn-off, we are also compelled to resort to binary evolution models. There are two ways of producing a blue straggler in a young open cluster. One is stable mass accretion, as detailed in the previous sections. The second one is the merger of a tight binary. If the merger takes place as a consequence of the expansion of the components during the hydrogen burning phase, it creates a star that is more massive than either of its progenitor stars, with a core hydrogen content that is higher than that of an equally old single star of the same mass \citep{schneider2015,wang2020}. If one of the stars that merged was close to the mass of the turn-off, the merger product will be a blue straggler that is brighter and bluer than the turn-off stars. Moreover, the kinematics of the system will not be altered.

In NGC~663, several objects deviate from the expected locations for single stars: the SBSs and the majority of BSGs. The fact that their kinematics are typical of cluster members may be thought as favouring the merger channel. However, as we are considering binary systems whose initial primaries had masses $\la11\:\mathrm{M}_{\sun}$, this is not necessarily the case. While isolated $11\:\mathrm{M}_{\sun}$ stars are expected to explode as supernovae, primaries in binaries of the same mass do not necessarily explode. For instance, \citet{siess2018} consider the evolution of a $11.2\:\mathrm{M}_{\sun}+9\:\mathrm{M}_{\sun}$ binary with different initial orbital configurations, finding final configurations that lead to merger, rejuvenated star plus ONe white dwarf, rejuvenated star + electron capture supernova or rejuvenated star + core collapse supernova, depending on the initial orbital period. It is therefore perfectly possible to envisage a system evolving into RX J0146.9+6121, while a very similar one evolves without a supernova explosion and ends up as a blue straggler + WD binary. 

The two mild SBSs located near the cluster centre, stars 4 and 30, have masses 11\,--\,13$\:\mathrm{M}_{\sun}$. They seem to require either a moderate amount of mass gain from a binary companion or have been produced by the merger of two stars well below the turn-off. The five high-luminosity BSGs, along with star 376, may well have been produced by a merger of stars close to the turn-off or through binary evolution that led to the formation of a WD from the core of the initial primary. Distinguishing between these formation channels is not easy with the current data.

If the original primary must be stripped down to a low-mass core to avoid explosion, He-rich material should be accreted. \citet{dray2007} suggested that accretion of such He-rich material would lead to evolution resembling that of stars of very high metallicity, which do not experience blue loops. On the other hand, \citet{menon24} found that roughly one third the BSGs in their LMC sample displayed He enhancement, which -- in combination with high N/C and N/O ratios -- they interpret as evidence for mergers. While theoretical expectations for both binary formation mechanisms imply a substantial He enrichment, we do not find evidence for significant He enhancement in any of the BSGs in NGC 663. In fact, within the large sample of \citet{deBurgos24}, no BSG of similar luminostiy displays significant He enhancement. BSGs in younger clusters, which must descend from progenitors that necessarily undergo core collapse, must be analysed to obtain a definite answer about the importance of the merger channel.

Nevertheless, we must stress a major difference between the predictions of binary evolution models and the observed population in NGC~663. In the majority of evolutionary channels that lead to rejuvenated stars, the accretor settles on a main-sequence track at higher luminosity compared to the original track because of the accretion of mass, and is therefore a main-sequence star close to critical rotation. Most of the stars that we see in NGC~663 requiring a binary history are evolved away from the main sequence. If they are formed via merger of a close binary, this would suggest that merger products rarely settle close to the ZAMS.

\section{Conclusions}

We have conducted a comprehensive investigation of NGC~663, the most prominent cluster in Cas OB~8 association and possibly in the whole Perseus arm. Our study integrates astrometric, photometric and spectroscopic analyses to characterise the cluster and its surroundings.

For the first time, we present optical spectroscopy for a large sample of stars in the region ($153$ stars, the vast majority being cluster members). We derive spectral types for all of them and stellar atmospheric parameters, $T_{\mathrm{eff}}$ and $\log\,g$, for over 40 of the brightest cluster members.

Using \textit{Gaia} EDR3 data, we identify $1456$ likely members based on proper motions and parallaxes, with a contamination rate below 2\% . The corrected median parallax ($0.37\pm0.05$) corresponds to a distance of $2.7$~kpc. Additionally, we identify a number of candidate runaway stars with parallaxes consistent with membership but divergent proper motions. 
Str\"{o}mgren  photometry reveals a moderate degree of differential reddening ($0.52$ to $0.79$), with an average value $E(b-y)$=$0.61\pm0.05$. The highest extinction values concentrate towards the northwestern region. From isochrone fitting, we obtain a  distance $2.7\pm0.4$~kpc, fully consistent with the  \textit{Gaia} value, and an age $22.5\pm2.5$ Ma. 

Our findings align with recent  \textit{Gaia} estimates to reveal  NGC~663 as potentially the most massive cluster in the Perseus arm, with NGC~7419 the only contender, amply surpassing in mass the whole area surrounding the double cluster h \& $\chi$Persei. There are more than $300$ astrometric B-type members, implying an initial mass likely exceeding $10^4\:\mathrm{M}_{\sun}$. NGC~663 contains many Be stars, almost 30 with spectra here, but the relative fraction to B stars (only $\sim30$\% above the turn-off and quickly falling off at lower masses) is not exceptionally high.

NGC~663 provides an excellent laboratory to study moderatley-massive stellar evolution. Our key findings in this respect are:

\begin{itemize}

\item Turn-off stars (with spectral types B2\,IV and B2.5\,IV) have masses somewhat above $8\:\mathrm{M}_{\sun}$, while giants extend between $9$ and $11\:\mathrm{M}_{\sun}$.

\item Four astrometric members and one star in the surrounding association have spectral types  earlier than B2 and can be considered spectroscopic blue stragglers. Among them, the Be/X-ray binary RX~J$0146.9+6121$ must have been formed through mass transfer. The original primary, with an initial mass around $11\:\mathrm{M}_{\sun}$, exploded as a supernova after transferring several solar masses of material to the progenitor of the current Be star. As a consequence, it is moving slowly away from the cluster. The remaining SBSs, which lack emission lines and exhibit typical cluster kinematics, are likely merger products.

\item The lack of Be blue stragglers, with the exception of the Be/X-ray binary, and the absence of Be runaways or walkways do not provide strong support for a binary formation channel, if they had to follow evolution similar to that of RX~J$0146.9+6121$. If case~A mass transfer from stars below the turn-off is frequent, they will leave remnants that are not massive enough to explode. Alternatively, we may have case~B mass transfer if we drop the assumption that stars of 10\,--\,$12\:\mathrm{M}_{\sun}$ starting their lives as primaries in binaries consistently undergo supernova explosions.

\item Five blue supergiant members and one blue straggler appear to have significantly higher masses (around $15\:\mathrm{M}_{\sun}$) than the brightest giants at the edge of the Hertzsprung gap (below $12\:\mathrm{M}_{\sun}$). Their typical cluster kinematics favour a formation mechanism via mergers, unless, again, we assume that initial primaries of 10\,--\,$12\:\mathrm{M}_{\sun}$ do not undergo supernova explosions.
\end{itemize}

Finally, Cas~OB8 remains a poorly studied association, despite its large size and very high mass. \textit{Gaia}  data enable a refined reassessment of its member clusters, revealing at least 18 associated clusters. Future studies should focus on precisely defining its extent and structure.

\section*{Data Availability}
 The data presented here are available based on reasonable requests to the authors.

\section*{Acknowledgements}

We thank the anonymous referee for corrections and helpful suggestions.

This research is partially supported by the Spanish
Government Ministerio de Ciencia, Innovaci\'on y Universidades and Agencia Estatal de Investigación (MCIU/AEI/10.130 39/501 100 011 033/FEDER, UE) under grants PID2021-122397NB-C21/C22.
AM and IN acknowledge the financial support of the MCIU with funding from the European Union NextGenerationEU
and Generalitat Valenciana in the call Programa de Planes Complementarios
de I+D+i (PRTR 2022), project HIAMAS, reference ASFAE/2022/017. NC acknowledges funding from the Deutsche Forschungsgemeinschaft (DFG) - CA 2551/1-2.

We thank Dr. Christian Motch and Dr. Olivier Hérent for obtaining the OHP observations.

Based on observations made with the Nordic Optical Telescope, operated
on the island of La Palma jointly by Denmark, Finland, Iceland,
Norway, and Sweden, in the Spanish Observatorio del Roque de los
Muchachos of the Instituto de Astrof\'{\i}sica de Canarias.  

Some of the data presented here have been taken using ALFOSC, which is owned by the Instituto de Astrofisica de Andalucia (IAA) and operated at the Nordic Optical Telescope under agreement between IAA and the NBIfAFG of the Astronomical Observatory of Copenhagen.

{\sc iraf} is distributed by the National Optical Astronomy Observatories, which are operated by the Association of Universities for Research in Astronomy, Inc., under cooperative agreement with the National Science Foundation

This research has made use of the SIMBAD, Vizier and Aladin
services developed at the Centre de Données Astronomiques de Strasbourg (CDS), France and of the WEBDA database, operated at the Department of Theoretical Physics and Astrophysics of the Masaryk University. We have made extensive use of the TOPCAT \citep{taylor2005} tool. 
This work has made use of data from the European Space Agency (ESA) mission Gaia
(https://www.cosmos.esa.int/gaia), processed by the Gaia Data Pro-
cessing and Analysis Consortium (DPAC, https://www.cosmos.esa.int/
web/gaia/dpac/consortium). Funding for the DPAC has been provided by
national institutions, in particular the institutions participating in the Gaia Multi-lateral Agreement. 








\newpage

\appendix

\section{Spectroscopic Tables}
\clearpage
\onecolumn

\begin{table*}
\begin{center}
	\caption{Stars with spectra observed with the INT at $R=4000$. Parallaxes have been zero point corrected.\label{tab:R4000}}
\begin{longtable}{lllcccc} 
	  \hline
          \noalign{\smallskip}
           RA&DEC&Spectral&WEBDA&Plx&pmRA&pmDE\\
	   (J2000)&(J2000)&Type&ID&(mas)&(mas$\cdot$a$^{-1}$)&(mas$\cdot$a$^{-1}$)\\
            \noalign{\smallskip}
\hline
\noalign{\smallskip}
\multicolumn{7}{c}{\textbf{Members}}\\
 \noalign{\smallskip}
\hline
\noalign{\smallskip}
01:45:53.73&$+$61:19:14.0&O9.2\,IV&162&0.382$\pm$0.016&$-$1.243$\pm$0.011&$-$0.572$\pm$0.012\\
\noalign{\smallskip}
01:47:00.19&$+$61:21:23.6&B0.7\,IVe&194&0.367$\pm$0.022&$-$1.029$\pm$0.016&$-$0.082$\pm$0.017\\ 
\noalign{\smallskip}
01:46:41.42&$+$61:13:15.2&B1\,III&30&0.356$\pm$0.017&$-$1.257$\pm$0.013&$-$0.292$\pm$0.014\\
\noalign{\smallskip}
01:46:38.99&$+$61:14:06.0&B1\,IV&4&0.336$\pm$0.016&$-$1.077$\pm$0.012&$-$0.450$\pm$0.013\\
\noalign{\smallskip}
01:46:33.96&$+$61:09:02.3&B2\,III&110&0.418$\pm$0.019&$-$1.228$\pm$0.014&$-$0.138$\pm$0.015\\
\hline
\end{longtable}
\end{center}
\justifying
{\footnotesize
$^6$ We consider this star as a member. See App.~\ref{app:runaways}.}
\end{table*}

\twocolumn

\clearpage

\onecolumn

\begin{longtable}{ccllccc}
	\caption{Stars with spectra observed with the INT only at $R=1400$. Parallaxes have been zero point corrected. \label{tab:R1400}}\\    
	  \hline
           \noalign{\smallskip}
           RA&DEC&Spectral &WEBDA&Plx&pmRA&pmDE\\	   
(J2000)&(J2000)&Type&ID&(mas)&(mas$\cdot$a$^{-1}$)&(mas$\cdot$a$^{-1}$)\\
            \noalign{\smallskip}
\hline
\endhead
 \multicolumn{7}{c}{\textbf{Members}}\\
 \hline
 \noalign{\smallskip}
01:46:45.67&+61:30:09.2&B0\,III&376&0.379$\pm$0.013&$-$1.267$\pm$0.009&$-$0.342$\pm$0.010\\
\noalign{\smallskip}
01:45:56.09&+61:12:45.2&B2\,IVe&93&0.358$\pm$0.017&$-$1.127$\pm$0.013&$-$0.414$\pm$0.014\\
01:47:26.76&+61:08:44.1&B2\,IVe&222&0.432$\pm$0.023&$-$1.279$\pm$0.017&$-$0.418$\pm$0.018\\
\noalign{\smallskip}
01:46:21.92&+61:13:27.1&B2\,V&18&0.397$\pm$0.020&$-$1.355$\pm$0.015&$-$0.291$\pm$0.017\\
01:46:57.76&+61:09:47.1&B2\,V&145&0.391$\pm$0.012&$-$1.201$\pm$0.008&$-$0.326$\pm$0.009\\
01:47:03.35&+61:07:43.1&B2\,V&184\footnote{This star has no values for parallax and proper motions, but we consider it a member because of its spectral type}&¨&¨&¨\\
\noalign{\smallskip}
01:45:46.39&+61:09:20.8&B2\,Ve&170&0.366$\pm$0.018&$-$1.274$\pm$0.014&$-$0.437$\pm$0.015\\
\end{longtable}

\twocolumn

\clearpage
\onecolumn

\begin{longtable}{ccllccc}
	\caption{Stars with spectra taken with the OHP ($R=900$) only. Parallaxes have been zero point corrected. \label{tab:R900}}\\
	  \hline
           \noalign{\smallskip}
           RA&DEC&Spectral&WEBDA&Plx&pmRA&pmDE\\
	   (J2000)&(J2000)&Type&&(mas)&(mas$\cdot$a$^{-1}$)&(mas$\cdot$a$^{-1}$)\\
            \noalign{\smallskip}
\hline
\endhead
\multicolumn{7}{c}{\textbf{Members}}\\
\hline 
\noalign{\smallskip}
01:46:07.87&$+$61:09:19.1&B2\,IV-V&135&0.420$\pm$0.016&$-$1.212$\pm$0.012&$-$0.595$\pm$0.013\\
\noalign{\smallskip}
01:46:30.62&$+$61:14:29.2&B2\,Ve&6&0.374$\pm$0.016&$-$1.143$\pm$0.011&$-$0.228$\pm$0.013\\
01:45:37.81&$+$61:07:59.1&B2\,Ve&210&0.362$\pm$0.014&$-$1.261$\pm$0.011&$-$0.366$\pm$0.011\\
\noalign{\smallskip}
01:43:05.07&$+$61:06:02.6&B2.5\,IV&160&0.373$\pm$0.030&$-$1.144$\pm$0.021&$-$0.503$\pm$0.022\\
\noalign{\smallskip}
01:45:49.06&$+$61:03:55.8&B2.5\,IV-V&247&0.376$\pm$0.017&$-$1.354$\pm$0.013&$-$0.393$\pm$0.014\\
\noalign{\smallskip}
01:46:50.67&$+$61:15:56.5&B2.5\,V&75&0.404$\pm$0.017&$-$1.206$\pm$0.012&$-$0.352$\pm$0.013\\
01:46:33.67&$+$61:16:47.7&B2.5\,V&78&0.358$\pm$0.013&$-$1.257$\pm$0.010&$-$0.331$\pm$0.011\\
\end{longtable}

\twocolumn

\begin{table*}
	\centering
	\caption{Atmospheric parameters for bright non-supergiant cluster members, derived from INT spectra. The values for $v\,\sin\,i$ are simply an estimate of line broadening, and should not be taken at face value. A lower limit of $50\:\mathrm{km}\,\mathrm{s}^{-1}$ is given for stars with no evident broadening.\label{tab:analysis}}
	\begin{tabular}{llccc} 
	  \hline
          \noalign{\smallskip}
          WEBDA&Spectral& $v\,\sin\,i$&$T_{\mathrm{eff}}$& $\log\,g$\\
	   ID&Type& ($\mathrm{km}\,\mathrm{s}^{-1}$)& (K)&\\
            \noalign{\smallskip}
\hline
\noalign{\smallskip}
\multicolumn{5}{c}{\textbf{Giants}}\\
 \noalign{\smallskip}
\hline
\noalign{\smallskip}
30	&B1\,III	& 50 & $24\,000 \pm	1500$  &  $3.8 \pm  0.2$\\
82	&B2\,III-IV & 70 & $17\,000 \pm	2000$  &  $3.2 \pm  0.3$\\
92	&B2.5\,IIIe & 130 &$17\,000 \pm 1800$   &  $3.1 \pm  0.2$\\
99	&B2.5\,III	& 150 & $18\,000 \pm	1000$  &  $3.1 \pm  0.1$\\
107	&B2.5\,III	& 130 & $19\,000 \pm	1600$  &  $3.2 \pm  0.2$\\
\hline\end{tabular}
\end{table*}

\clearpage
\onecolumn
\section{Photometric Tables}
\begin{landscape}
\begin{longtable}{lllrrrrrrrrrrrcccc}
\caption{Coordinates in J2000, $uvby$$\beta$ photometry and spectral types for Gaia members.  Material on-line\label{tab:membersphot}}\\
\hline
RA(J2000)& DEC(J2000)&$V$&$\sigma_{V}$&$N_{V}$&$(b-y)$&$\sigma_{b-y}$&$N_{b-y}$&$m_{1}$&$\sigma_{m_{1}}$&$N_{m_{1}}$&$c_{1}$&$\sigma_{c_{1}}$&$N_{c_{1}}$&$\beta$&$\sigma_{\beta}$&$N_{\beta}$&SpT\\
\hline\hline
\endhead
\hline

01:45:57.69 &  $+$61:09:13.8&   13.600&   0.022&   3  &  0.483&   0.019&   3 &  $-$0.022 &  0.028 &  3  & 0.352 &  0.027 &  2	 & 2.710  &  0.019  & 3     &  B4\,V	 \\	       
01:46:59.28 &  $+$61:09:15.0&   15.026&   0.043&   2  &  0.653&   0.016&   2 &  $-$0.068 &  0.001 &  2  & 0.940  &  0.015 &  2	 & -	 &  -     & 0 &   -	 \\    
01:46:25.26 &  $+$61:09:27.5&   15.544&   0.004&   2  &  0.526&   0.015&   2 &  0.033	 &  0.017 &  2  & 1.036 &  0.060  &  1	 & 2.880  &  0.050   & 1     & - \\
01:46:38.50 &  $+$61:09:30.9&   13.630&   0.006&   3  &  0.485&   0.017&   3 &  $-$0.023 &  0.028 &  3  & 0.498 &  0.039 &  3	 & 2.708 &  0.039  & 3     &  B5\,Vp?	 \\    
01:46:51.11 &  $+$61:09:31.5&   15.067&   0.005&   2  &  0.564&   0.026&   2 &  $-$0.007 &  0.048 &  2  & 0.931 &  0.021 &  2	 & 2.839 &  0.044  & 2     &  -	 \\
01:46:38.68 &  $+$61:09:43.3&   13.474&   0.006&   3  &  0.490&   0.017&   3 &  $-$0.001 &  0.038 &  3  & 0.383 &  0.021 &  3	 & 2.707 &  0.015  & 3     &B4\,V	 \\  
01:46:21.90 &  $+$61:09:45.7&   15.144&   0.008&   3  &  0.526&   0.018&   3 &  0.020	 &  0.051 &  3  & 0.876 &  0.006 &  3	 & 2.840  &  0.027  & 3     & -	 \\
01:46:16.57 &  $+$61:09:46.8&   14.477&   0.012&   3  &  0.548&   0.002&   3 &  $-$0.012 &  0.004 &  3  & 0.694 &  0.044 &  3	 & 2.755 &  0.027  & 3     & -	 \\
\hline
\end{longtable}
 \end{landscape}

\onecolumn

\begin{longtable}{llccrcr}
\caption{Coordinates in J2000 and photometric values in $E(b-y)$, $Ec_{1}$, $c_{0}$, $V_{0}$ and $M_{V}$ for Gaia members in NGC~663. Material on-line\label{tab:abs}}\\
\hline
RA(J2000)& DEC(J2000)&$E(b-y)$&$Ec_{1}$&$c_{0}$&$V_{0}$&$M_{V}$\\
\hline\hline
\endhead
\hline
01:45:57.69  &  $+$61:09:13.8   &    0.587 &  0.117 &  0.235 &  11.076 & $-$1.12  \\
01:46:59.28  &  $+$61:09:15.0   &    0.704 &  0.141 &  0.799 &  11.997 & $-$0.20  \\
01:46:25.26  &  $+$61:09:27.5   &    0.563 &  0.113 &  0.923 &  13.124 & 0.92     \\
01:46:38.50  &  $+$61:09:30.9   &    0.574 &  0.115 &  0.383 &  11.160 & $-$1.04  \\
01:46:51.11  &  $+$61:09:31.5   &    0.613 &  0.123 &  0.808 &  12.432 & 0.23     \\
01:46:38.68  &  $+$61:09:43.3   &    0.591 &  0.118 &  0.265 &  10.932 & $-$1.27  \\
01:46:21.90  &  $+$61:09:45.7   &    0.579 &  0.116 &  0.760 &  12.655 & 0.45     \\
01:46:16.57  &  $+$61:09:46.8   &    0.620 &  0.124 &  0.570 &  11.811 & $-$0.39  \\
\hline
\end{longtable}
 
\twocolumn 
\clearpage
\section{Extra figures}
\clearpage

\begin{figure*}
\centering
\resizebox{\textwidth}{!}{\includegraphics[]{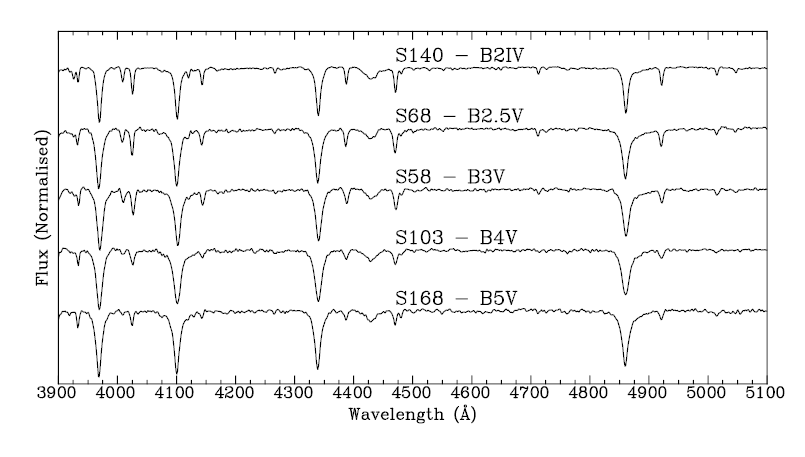}}
\caption{Stars in NGC~663 observed with IDS + R400V. The spectra have $R\approx1\,400$ and varying SNR. Although the exposure time was increased for fainter stars, their SNR is generally lower, because of the very large difference in magnitude. At this resolution, spectral classification is given by the He\,{\sc i}~4471/Mg\,{\sc 4481} ratio and the decreasing strength of He\,{\sc i} lines with increasing spectral type. The C\,{\sc ii}~4267\,\AA\ is seen in the earliest objects, as well as other very weak metallic lines\label{R400sample}.}
\end{figure*} 

\begin{figure*}
\centering
\resizebox{\textwidth}{!}{\includegraphics[]{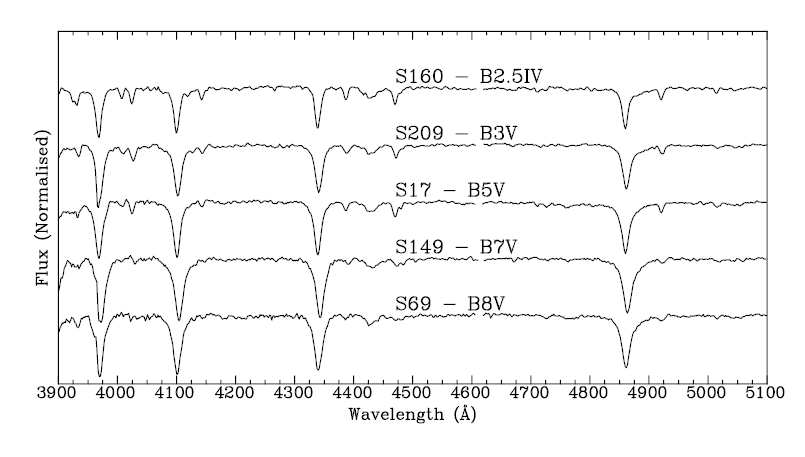}}
\caption{Stars in NGC~663 observed with \textit{Carelec}. The spectra have $R\approx900$ and varying SNR, generally lower for later types. Classification criteria are as for the IDS + R400V sample. There is a gap around 4615\,\AA\ in all the spectra due to a defect in the CCD\label{OHPsample}.}
\end{figure*} 

\clearpage
\section{Be star census}
\label{app:bes}

\citet{sanduleak1979} provided a list of Be stars in NGC~663, which he later updated after removing some objects that had probably been misidentified \citep{sanduleak1990}. We have observed
all the 29 objects in the first list, except for 297 = San~27, which is very
far away from the cluster centre, but is a member according to \citet{hunt2023}. None of the objects that Sanduleak removed from the list of Be stars in the second paper -- namely, 30 = San 18, 67 = San 19, 91 = San 7, 110 = San 15 = BD~$+60\degree$\,344, and 147 = San 24 = BD~$+60\degree$\,347 -- shows any trace of emission in our spectra. In particular, 147 is an B9\,Ib supergiant and cannot be, by definition, a Be star. Additionally, 107 = San~11 = BD~$+60\degree$\,340 was marked by \citet{sanduleak1979} as a very likely wrong identification for MWC~700, as it has not shown any emission at least since 1946. It is quite possible that MWC~700 must be identified with 141 = San~13 = VES~619, instead. 
We detect all other 23 confirmed Be stars, with the following caveats:

\begin{itemize}

\item The brighter component of BD~$+60^{\degree}$\,332, 92 = San~6 = BD~$+60^{\degree}$\,332A is a Be star, while 91 = San~7 = BD~$+60^{\degree}$\,332B is a
normal star. The latter had already been removed by \citet{sanduleak1990}.
In the WEBDA, 91 is given as the brighter component, but
from our spectra, it is the star 
identified in both Simbad and the WEBDA as 92 which is brighter and a
Be star, in agreement with Sanduleak.

\item In the WEBDA, San 13 = BD $+60\degree$\,341 is
identified with 144. This is not a Be star. The correct
identification is the rather brighter 141.

\item WEBDA star 7191 does not correspond to any star. BD~$+60\degree$\,333 has only two components. BD~$+60\degree$\,333A = 54 is a supergiant, while BD~$+60\degree$\,333B = 53 = San 8 is a Be star.

\item Star 40 does not correspond with any real object. It is an old (offset) position for BD~$+60\degree$\,343. The fainter component, BD~$+60\degree$\,343B =
8072 = San 16, is a Be star, while the brighter component, BD~$+60\degree$\,343A = 830, is a supergiant. Star 39 is another
misidentification for the same double or one of its components. \citet{sanduleak1990} states that no emission has been visible since the
1950's, but 8072 is a rather strong Be star in our spectra.

\item In the WEBDA, Star 120 is given as the identification of VES 624 = San 21, but this is a foreground late-type star. The correct identification is
its neighbour, 121.
\end{itemize}

After Sanduleak's work, different authors have searched for Be stars in the cluster by resorting to different methods.

\citet{capilla2000}, using H$\alpha$ and
H$\beta$ filters, identify stars 61, 102 and 175 as candidate Be stars. We confirm 61, but we do not find any sign of emission in either 102 or 175, which were
not detected by either \citet{pigulski2001} or \citet{yu2015}.

\citet{pigulski2001}, using H$\alpha$ filters, identify three new
candidate Be stars. We confirm that stars 128 and 151 are emission-line B
stars. Star 240, on the other hand, is a foreground G-type star.
 
\citet{yu2015} found four further emission-line candidates. Of these, three are faint and not astrometric members, while one (UCAC4 755$-$019877) is outside the field surveyed, about $23\arcmin$ to the SW of the cluster. Its proper motions are not incompatible with ejection from NGC~663.

In this work, we find 3 new objects showing clear, if weak, emission in H$\alpha$, namely stars 101, 132 and 181. Additionally, 131 shows considerable emission infilling, not quite reaching the continuum level, and would not be classified Be from the blue spectrum alone. Finally, 97 shows asymmetric Balmer  absorption, likely to be filled in by emission, though the case of a double-lined spectroscopic binary cannot be ruled out at this resolution.

Therefore, we confirm via spectroscopy 26 Be stars previously reported and find three new ones, plus two candidates.

\clearpage
\section{Analysis of astrometric nonmembers for possible runaway stars}
\label{app:runaways}

Several stars observed spectroscopically are not astrometric members of NGC~663, despite having spectral types and photometry compatible with membership, and parallaxes not very different from those of cluster members. Here we address the possibility that these stars are connected to the cluster, by considering two hypothesis. 

First, a few stars have proper motions compatible with membership, but parallaxes outside the adopted range. However,  \textit{Gaia} EDR3 parallax uncertainties represent only (random) internal errors and do not reflect the true dispersion of catalogue values \citep{maiz2021}. To address this, we applied the method from \citet{maiz2021} to derive more accurate external errors and reassessed whether the revised error bars allow membership classification.

Second, a larger number of stars have parallaxes consistent with membership, but exhibit proper motions outside the range considered for members. For these cases, we calculated relative proper motions with respect to the cluster by subtracting the median values listed in Table~\ref{tab:median} from their proper motions. We then examined whether these relative proper motions are consistent with ejection from the cluster.

Based on this analysis, we reached the following conclusions:

\begin{itemize}

\item Table~\ref{tab:R900}: {\bf Star 183} has proper motions in the range for members. We re-evaluate the parallax error by using the procedure of \citet{maiz2021}, finding a value $0.021\:$mas$\cdot$a$^{-1}$ that makes the parallax compatible with membership. {\bf Star 237} lies very slightly outside the membership range in both parallax and pmRA (this by only $0.05\:$mas$\cdot$a$^{-1}$). With an external parallax error of $0.022\:$mas$\cdot$a$^{-1}$, both values are nearly compatible at the 1--$\sigma$ level and we therefore take it as a likely cluster member. Among non-members,  {\bf Star 153} lies about $9\arcmin$ to the NE of the centre of the cluster. Its relative proper motions ($-0.162$ mas$\cdot$a$^{-1}$ , $-0.682$ mas$\cdot$a$^{-1}$) imply that it moves towards the cluster; this is thus an association star rather than a member. 

\item Table~\ref{tab:R1400}: {\bf Star 20} is in the center of the cluster where the density of stars is high. Its relative proper motions ($-0.458$ mas$\cdot$a$^{-1}$ , $-0.153$ mas$\cdot$a$^{-1}$) indicate motion in the SW direction. We consider this star as a member of the cluster that has been dynamically perturbed by interactions with other stars. {\bf Star 239} lies $8.5\arcmin$ to the W of the cluster centre. Its relative proper motions ($0.223$ mas$\cdot$a$^{-1}$ , $-0.574$ mas$\cdot$a$^{-1}$) imply a motion unrelated to the cluster; this is an association star.  {\bf Star 168} lies to the W and somewhat to the S of the cluster centre, at $\sim5\arcmin$. Its relative proper motions ($0.3$ mas$\cdot$a$^{-1}$ , $-0.6$ mas$\cdot$a$^{-1}$) suggest that it must have originated outside the cluster. {\bf Star 117} has a slightly higher parallax than the highest value allowed, but its external error is 0.035 mas$\cdot$a$^{-1}$, meaning that it is almost compatible with that of the cluster. This star is to the SE of the centre ($\sim5\arcmin$) and its relative proper motions are ($0.706$ mas$\cdot$a$^{-1}$ , $-0.004$ mas$\cdot$a$^{-1}$), suggesting that it may have been ejected from the cluster.
 
\item Table~\ref{tab:R4000}: {\bf Star 91} lies close to the center of the cluster. Its relative proper motions ($0.007$ mas$\cdot$a$^{-1}$ , $-0.476$ mas$\cdot$a$^{-1}$) indicate motion towards the S. We assume that this is a dynamically perturbed cluster member. {\bf Star 129} lies within the cluster, to the NW of the centre, in an area of low density. Its relative proper motions ($0.045$ mas$\cdot$a$^{-1}$ , $-0.614$ mas$\cdot$a$^{-1}$) imply a motion towards the S, implying that it is likely unrelated to the cluster.  {\bf Star 288} is the only Be star whose proper motions indicate definite non-membership. Located $\sim13\arcmin$ to the S and slightly to the E of the cluster core, its relative proper motions ($-0.016$ mas$\cdot$a$^{-1}$ , $-0.421$ mas$\cdot$a$^{-1}$) indicate motion to the S. This star is a possible ejection, if it originates from the eastern extension of the cluster core. 

\item  Table~\ref{tab:SGFIES}: {\bf Star 307} lies $\sim14\arcmin$ to the N and somewhat to the W of the cluster core. Its relative proper motions ($-0.131$ mas$\cdot$a$^{-1}$ , $0.412$ mas$\cdot$a$^{-1}$) imply motion in the NW direction, suggesting that this A0\,Ib supergiant may have been ejected from the cluster, most likely from the Eastern extension to the core.

\end{itemize}

\clearpage
\section{Members of Cas~OB8}

\begin{table*}
	\centering
	\caption{Median values in EDR3 of parallax and proper motions for clusters in Fig.~\ref{CASOB8}. Names in bold are the likely members of the Cas OB8 association.  \label{tab:clusters_cas0b8}}
        \begin{tabular}{lcrcrrc}
       Name&Longitude&Latitude&Plx&pmRA&pmDE&log\it t\\
       &(\degree)&(\degree)&(mas)&(mas$\cdot$a$^{-1}$)&(mas$\cdot$a$^{-1}$)&(years)\\
        \hline
    
    FSR~0534&126.05&-1.42&$0.32\pm0.04$&$-0.36\pm0.05$&$-0.42\pm0.05$&7.28\\
    HSC~1014&126.31&-0.17&$0.29\pm0.03$&$-1.58\pm0.07$&$-0.18\pm0.05$&7.14\\
    NGC~457&126.63&-4.39&$0.33\pm0.03$&$-1.59\pm0.07$&$-0.73\pm0.08$&7.36\\
    HSC~1019&127.07&0.13&$0.32\pm0.02$&$-1.61\pm0.03$&$-0.24\pm0.04$&7.88\\
    OC~0237&127.45&-0.11&$0.34\pm0.03$&$-1.59\pm0.05$&$-0.41\pm0.05$&7.35\\
   \textbf{HSC~1025}&128.01&-0.55&$0.32\pm0.01$&$-1.67\pm0.04$&$-0.25\pm0.04$&7.61\\
   \textbf{NGC~581}&128.06&-1.80&$0.37\pm0.03$&$-1.39\pm0.07$&$-0.60\pm0.06$&7.44\\
  \textbf{HSC~1026}&128.08&0.71&$0.34\pm0.03$&$-1.42\pm0.03$&$-0.48\pm0.04$&7.67\\
    UBC~1605&128.13&1.66&$0.31\pm0.02$&$-2.04\pm0.05$&$0.10\pm0.05$&7.55\\
    UBC~41&128.15&-2.67&$0.41\pm0.04$&$-0.75\pm0.07$&$-0.83\pm0.07$&7.92\\
   \textbf{Trumpler~1}&128.22&-1.13&$0.36\pm0.03$&$-1.65\pm0.09$&$-0.58\pm0.07$&7.44\\
   \textbf{NGC~637}&128.55&1.73&$0.34\pm0.03$&$-1.26\pm0.07$&$-0.04\pm0.07$&7.30\\
   \textbf{UBC~602}&128.70&-1.82&$0.34\pm0.02$&$-1.22\pm0.08$&$-0.52\pm0.05$&7.61\\
    \textbf{NGC~654}&129.08&-0.35&$0.33\pm0.05$&$-1.15\pm0.09$&$-0.33\pm0.09$&7.28\\
   \textbf{HSC~1043}&129.36&-4.84&$0.33\pm0.03$&$-1.03\pm0.07$&$-0.73\pm0.10$&7.96\\
   \textbf{NGC~659}&129.37&-1.54&$0.30\pm0.04$&$-0.84\pm0.08$&$-0.32\pm0.07$&7.62\\
   \textbf{NGC~663}&129.47&-0.98&$0.35\pm0.03$&$-1.14\pm0.08$&$-0.34\pm0.09$&7.44\\
   \textbf{Berkeley~6}&130.10&-0.98&$0.32\pm0.03$&$-0.90\pm0.08$&$-0.53\pm0.07$&7.94\\
    \textbf{Berkeley~7}&130.13&0.37&$0.34\pm0.03$&$-1.02\pm0.11$&$-0.25\pm0.07$&7.56\\
   \textbf{HSC~1055}&130.35&$-5.99$&$0.32\pm0.04$&$-0.90\pm0.08$&$-1.01\pm0.07$&7.89\\
  \textbf{UBC~1229}&130.49&$-2.89$&$0.33\pm0.02$&$-0.70\pm0.09$&$-0.42\pm0.07$&7.80\\
    Czernick~5&130.55&$-0.56$&$0.27\pm0.08$&$-2.01\pm0.10$&$-0.37\pm0.08$&7.81\\
    HSC~1059&130.66&1.82&$0.29\pm0.03$&$-2.06\pm0.09$&$-0.63\pm0.05$&7.78\\
   \textbf{HSC~1060}&130.67&0.07&$0.34\pm0.02$&$-1.02\pm0.06$&$-0.39\pm0.09$&7.38\\
   \textbf{Czernick~6}&130.90&1.07&$0.33\pm0.03$&$-1.19\pm0.06$&$-0.20\pm0.06$&7.63\\
  \textbf{UBC~604}&130.96&3.78&$0.34\pm0.02$&$-0.74\pm0.05$&$0.41\pm0.04$&7.53\\
    HSC~1066&131.32&-0.24&$0.29\pm0.05$&$0.34\pm0.10$&$-0.45\pm0.07$&7.78\\
  \textbf{UBC~1232}&131.39&2.58&$0.31\pm0.03$&$-0.69\pm0.05$&$0.27\pm0.07$&7.63\\
    UBC~1234&132.10&-2.70&$0.40\pm0.02$&$-1.17\pm0.06$&$-0.85\pm0.04$&7.65\\
    Riddle~4&132.22&-1.23&$0.35\pm0.03$&$-0.81\pm0.05$&$-0.50\pm0.06$&7.16\\
    COIN-Gaia~35&133.06&$-$1.21&$0.37\pm0.02$&$-0.55\pm0.04$&$-0.46\pm0.05$&7.73\\
    UBC~86&133.59&-3.53&$0.38\pm0.03$&$-0.84\pm0.07$&$-1.01\pm0.04$&7.64\\
    UBC~418&133.75&-1.41&$0.35\pm0.02$&$-0.50\pm0.08$&$-0.52\pm0.06$&7.64\\
    OC~0243&133.78&2.15&$0.32\pm0.02$&$-0.32\pm0.11$&$-0.15\pm0.10$&7.44\\

        \hline
        \end{tabular}
\end{table*}


\bsp	
\label{lastpage}
\end{document}